\documentclass[preprint,aps,epsfig]{revtex4}
\usepackage{amsfonts}
\usepackage{amssymb}
\usepackage{graphics,psfig,graphicx}

\newcommand {\nb}{\nonumber\\}

\newcommand {\be}{\begin{equation}}
\newcommand {\ee}{\end{equation}}
\newcommand {\bea}{\begin{eqnarray}}
\newcommand {\ea}{\end{eqnarray*}}
\newcommand {\ba}{\begin{eqnarray*}}
\newcommand {\eea}{\end{eqnarray}}
\newcommand {\ham} {{\mathcal H}}
\newcommand {\lleb} {{\mathcal L}}
\newcommand {\n} {{\mathcal N}}
\newcommand {\mdvr} {{\mathcal U}}
\newcommand {\lscat} {{\mathcal L}}
\newcommand {\bra}{\langle}
\newcommand {\ket}{\rangle} 
\newcommand {\rv}{{\bf r}}

\newcommand {\yv}{{\bf y}}
\newcommand {\xv}{{\bf x}}

\newcommand {\refeq}[1] {(\ref{#1})}
\begin{document}

\title{Three-Nucleon Continuum by means of the Hyperspherical Adiabatic Method }
\author{Paolo Barletta}
\affiliation{Department of Physics and Astronomy, University College London, Gower Street, WC1E6BT London, UK}
\author{A. Kievsky}
\affiliation{Istituto Nazionale di Fisica Nucleare, Piazza Torricelli 2, 56100 Pisa, Italy}
%\date{\today}

\begin{abstract}

This paper investigates the possible use of the Hyperspherical Adiabatic
 basis in the description of scattering states of a three-body system.
 In particular, we analyze a 1+2 collision process below the three-body
 breakup. The convergence patterns for the observables of interest are
 analyzed by comparison to a unitary equivalent Hyperspherical Harmonic
 expansion. Furthermore, we compare and discuss two different possible
 choices for describing the asymptotic configurations of the system,
 related to the use of Jacobi or hyperspherical coordinates.
 In order to illustrate the difficulties and advantages of
 the approach two simple numerical applications are shown in the case of
 neutron-deuteron scattering at low energies using $s$-wave interactions.
We found that the
 optimization driven by the Hyperspherical Adiabatic basis is not as
 efficient for scattering states as in bound state applications.
\end{abstract}

\maketitle

\section{Introduction}

The Hyperspherical Adiabatic (HA) method is based on the parametrization of
the internal degrees of freedom with hyperspherical coordinates 
(see Refs.\cite{nielsen} and references therein).
The method then
consists in expanding the system's wavefunction on a basis made of 
hyperangular optimized functions (the adiabatic basis set) 
times (unknown) hyperradial functions.
The hyperangular basis elements are taken as the eigenvectors of the  
Hamiltonian operator for a fixed value of the hyperradius $\rho$. 
Once those eigenvectors have been calculated, the hyperradial functions are
obtained as the solutions of a system of coupled one-dimensional
differential equations. 
The advantages of such approach
are that the HA basis should drive a quick convergence for the expansion, due 
to its optimization, the payback is represented by the necessity and the
difficulty in calculating accurately the first and second derivatives of the
adiabatic basis set with respect  to the hyperradius. Those terms are crucial
to the method as they represent the coupling terms between the various
hyperradial  differential equations. In some applications of the HA method it 
was shown that the strong coupling between pair of elements of the adiabatic
basis makes the hyperradial problem particularly hard to solve \cite{bge00}.

The properties of the adiabatic basis functions have been object of several 
studies and are well-known. In particular, in the asymptotic limit of large 
hyperradius the HA functions are known to converge towards the scattering 
states of the three-body system, both below and above break-up.
This characteristic makes the adiabatic expansion a valid choice
to describe the three-body continuum states. 
In the literature there are several studies of the bound spectrum 
of a three-nucleon system by means of the HA method \cite{dcf82,bfr82,ffs88},
but very few dealing with continuum states \cite{fab1}.
This paper investigates the possibility of using
the HA approach to describe a three-body elastic process in which
a particle collides the other two, initially forming a bound state.
The object of this work is the study of the appropriate boundary conditions 
to be imposed to  the hyperradial functions as $\rho\rightarrow\infty$ and 
a careful analysis of the convergence properties of the HA expansion.

In order to quantitatively understand the pattern of convergence of 
the HA expansion we make use of the parallelism that can be built 
between the HA method and the Hyperspherical Harmonic (HH) expansion.
In fact, we can consider two different expansions for the system's 
wavefunction, one in terms of $N_A$ HA basis elements, and the second 
in terms of $N_H$ HH basis elements. 
When $N_A=N_H$ the two expansions are connected by 
a unitary transformation and therefore must yield identical results.
Since the HH basis has been used several times to describe
scattering states~\cite{krv94,kvr97}, we exploit this knowledge to study
the convergence of the HA expansion.
In particular, we will study the
convergence properties of the $L=0$ phase shift at low energies in 
a $1+2$ collision, which has been used as a benchmark problem 
in literature (see for example~\cite{cpf89}). 

The problem of the boundary conditions to be imposed to the 
hyperradial functions is related to the difficulties associated 
with obtaining the eigenvectors and eigenvalues of the adiabatic Hamiltonian 
at large values of the hyperradius. As the lowest adiabatic functions 
tend to the two-body bound wavefunctions, an accurate description 
of those states using, for example, the expansion in HH functions is
known to be very difficult. This is because, as $\rho\rightarrow\infty$,
the two-body bound states are localized in a very small zone of the
hyperangular phase-space. Consequently, this particular configuration
necessitates a large number of HH functions to be described \cite{bge00}.
In fact, it can be shown that the number of HH required to reproduce 
this type of spatial configuration grows exponentially with the hyperradius.
If the interest is limited to study deep three-body
bound states, the problem just described does not manifest and 
a tractable number of HH functions suffices for a good accuracy.
Due to the finite hyperradial size of the associated wavefunction,
the adiabatic Hamiltonian needs to be solved only up to a non so
large value of the hyperradius. 
However, there are cases in which shallow bound states are present (as
Efimov states) and the adiabatic Hamiltonian needs to be solved for
very large values of $\rho$. Furthermore, for energies in the continuum,
the associated three-body scattering wavefunction has an infinite
extension and a direct application of the HA necessitates of the
solution of the adiabatic Hamiltonian at very large values of $\rho$,
too. In order to obtain accurate asymptotic solutions to the adiabatic 
Hamiltonian we have followed in detail the procedure outlined by 
Nielsen and co-workers~\cite{nielsen}.

Finally, interest in this work is also sparkled by an article of Fabre de la 
Ripelle~\cite{fab1}, where he suggested the possibility of expanding
the three-body 
asymptotic scattering states into the adiabatic basis set, and retaining only 
the first term in such an expansion, resulting in a considerable reduction
of the numerical burden. We will analyze this truncation together with
the contribution of higher terms.

This paper is organized as follows: in the next section the HA method is 
presented, by first introducing the notation. The expansion of the HA
basis in terms of the HH functions is given as well as the method
to describe the HA functions and the adiabatic potentials at large
values of $\rho$.
Section III treats the problem of scattering states. Two different 
methods of implementing the Kohn Variational Principle are given 
in conjunction with the HA basis. The asymptotic conditions
are given in terms of the distance between the incident particle and
the two-body system and in terms of $\rho$. Section IV is devoted to numerical
applications. Results are presented using a simple Gaussian two-body 
potential and the semi-phenomenological $s$-wave MT-III potential~\cite{pfg82}.
The final Section is devoted to the conclusions and perspectives.

\section{Hyperspherical Adiabatic Method}

Let us consider a system of three identical particles of 
mass $m$, in a state of total orbital angular momentum $L=0$. 
Other quantum
numbers are represented by the total spin $S$, total isospin $T$, 
and the symmetry under particle 
permutation $\Pi$, which can take the values $a$
(anti-symmetric, for three fermions) or $s$
(symmetric, in the case of three bosons).
A further quantum number needed to uniquely identify each
wavefunction is given by the vibrational number
$n$ ($n=1,2, ...$) for bound states or the energy $E$ for continuum states.

Let us start from the definition of Jacobi coordinates $\{\xv_i,\yv_i\}$ 
\bea
\xv_i & = & {1\over \sqrt{2}} (\rv_j - \rv_k)  \nb
\yv_i & = & {1\over \sqrt{6}} (\rv_j + \rv_k - 2 \rv_i)
\label{jacs}
\eea
where $\{ \rv_i \}$ are the Cartesian coordinates of the three particles and
$i,j,k=1,2,3$ cyclic. The hyperspherical variables 
$\{\rho,\theta_i\}$ are defined as follows
\be
  X_i=\rho\cos\theta_i, \quad Y_i=\rho\sin\theta_i
\label{hangle}
\ee
where $\rho$ is the hyperradius which is symmetric under any permutation of 
the three particles and $\theta_i$ is the hyperangle, which is
 dependent on the 
particular choice of the Jacobi coordinate system. In terms
of the interparticle distances $r_{ij}=|\rv_i-\rv_j|=\sqrt{2} X_k$ the hyperradius reads:
\be
  \rho={1\over \sqrt{3}}\sqrt{r^2_{12}+r^2_{23}+r^2_{31}} \ .
\ee
In addition to $\rho$ and $\theta_i$ there are four more
coordinates needed to parametrize all the possible spacial 
configurations of the three 
particles, for example the four polar angles which define 
the orientation of the two Jacobi vectors with respect to
 the laboratory frame of reference.
However, in the particular case of total orbital angular momentum $L=0$, the number of such coordinates  
can be reduced to just one non-trivial functional dependence, represented by the 
cosine $\mu_i$ of the angle between  the two Jacobi vectors $\{\xv_i,\yv_i\}$:
\be
\mu_i = \xv_i \cdot \yv_i / (X_i Y_i) .
\ee
In the following we will refer to the set of hyperangles $\{\theta_i,\mu_i\}$
as $\Omega_i$, or more in general as $\Omega=\{\theta,\mu\}$ when there is 
no need to specify the choice of a particular permutation of the particles
defining a set of Jacobi coordinates.

The Hamiltonian operator $\ham$ takes the following 
expression in hyperspherical coordinates
\be
\ham =  -\frac{\hbar^2}{2 m} T_\rho + \frac{\hbar^2}{2 m \rho^2}G^2 + V(\rho,\Omega)   ,
\label{hhc}
\ee
where $V$ is the potential energy operator, $T_\rho$ is the hyperradial operator
\be
T_\rho =  \frac{d^2}{d \rho^2} + \frac{5}{\rho} \frac{d}{d\rho}
\label{trho}
\ee
and $G^2$ is the grand-angular operator 
\be
G^2 = \frac{4}{\sqrt{1-z^2}}\frac{d}{dz}(1-z^2)^{3/2}\frac{d}{dz} +\frac{\ell_x^2}{\cos^2\theta} +\frac{\ell_y^2}{\sin^2\theta} . 
\label{gqua}
\ee
where $z=\cos 2\theta$ and $\ell_x$ and $\ell_y$ are the angular momentum operators associated 
with the $\xv$ and $\yv$ vectors, respectively. 
The volume element is $\rho^5 d\rho \sqrt{1-z^2}dz d\mu$.

The system wavefunction $\Psi$, with quantum numbers $L$, $S$, $T$, $\Pi$, and $n$ (or $E$), is expanded as follows:
\be 
\Psi^{LST\Pi}_n = \sum_{\nu=1}^{\infty} u^{n}_\nu(\rho) \Phi^{LST\Pi}_\nu(\rho,\Omega),
\label{adbasis}
\ee
where $\{\Phi^{LST\Pi}_\nu\}$ are the eigenfunctions of the operator $\ham_\Omega$ made of the hyperangular part of the kinetic operator plus the potential energy operator, in which $\rho$ acts only as a parameter:
\be
\ham_\Omega \Phi^{LST\Pi}_\nu = \left[ \frac{\hbar^2}{2 m \rho^2}G^2 + V \right] \Phi^{LST\Pi}_\nu(\rho,\Omega) = U_\nu(\rho) \Phi_\nu^{LS\Pi}(\rho,\Omega).
\label{adeq}
\ee
The set of eigenfunctions $\{\Phi_\nu^{LS\Pi}\}$ is known as the adiabatic basis set, and the 
associated eigenvalues $\{ U_\nu(\rho)\}$ as the adiabatic curves or potentials. In practical 
calculations, the infinite expansion of eq. \refeq{adbasis} needs to be truncated to a finite 
number of basis elements, say $N_A$. The convergence for the observables of interest with 
respect to this parameter is then checked.

The initial Hamiltonian problem is thus tackled in two steps:
firstly, the HA basis functions $\{\Phi^{LS\Pi}_\nu\}$
and the associated potentials  $\{U_\nu(\rho)\}$ are calculated by 
solving eq. \refeq{adeq}. Secondly,
the hyperradial functions $u^{n}_\nu(\rho)$ are obtained 
as the solutions of a system of $N_A$
coupled one-dimensional differential equations, which can be expressed as follows
 \cite{gpk90}:
\bea
\sum_{\nu=1}^{N_A} & & \left[ \left(-\frac{\hbar^2}{2 m}T_\rho 
+U_\nu-E\right) \delta_{\nu'\nu} + B_{\nu'\nu} \right] u_\nu
+  C_{\nu'\nu}\frac{d}{d\rho}u_\nu \nb
&+& \frac{d}{d\rho}\left( C_{\nu'\nu} u_\nu \right) =0 \ \ \ (\nu'=1,\dots,N_A),
\label{usys}
\eea
where the coupling terms $B_{\nu'\nu},C_{\nu'\nu}$ follow from the dependence on $\rho$ 
of the HA basis  :
\be
B_{\nu'\nu}(\rho) = \frac{\hbar^2}{m \rho^2} \bra \frac{d\Phi_{\nu'}}{d\rho} 
| \frac{d \Phi_\nu}{d\rho} \ket_\Omega,
\ee
and 
\be
C_{\nu'\nu}(\rho) = \frac{\hbar^2}{m \rho^2} \bra \Phi_{\nu'} | 
\frac{d \Phi_\nu}{d\rho} \ket_\Omega. \ee 
 For bound states 
solutions, and short range potentials, the functions $\{ u_\nu \}$ tend 
to zero exponentially as $\rho \rightarrow \infty$, whereas for scattering 
states the boundary conditions to be imposed to the $\{ u_\nu \}$ will 
be discussed in the next Section.

The first step in the implementation of an HA calculation consists in obtaining
the adiabatic basis elements and the associated adiabatic potentials,
 solutions of eq. \refeq{adeq}, for a number of values of $\rho$.
Among several available techniques we have chosen to use a variational 
approach, by expanding the functions $\{\Phi^{LST\Pi}_\nu\}$ onto a set of 
Hyperspherical Harmonics (HH) of size $N_H$. In order to define a basis set 
with the desired properties under particle permutation, we combine opportunely 
hyperspherical polynomials based on different set of Jacobi coordinates 
\cite{kvr97}. 
The expansion for $\Phi^{LST\Pi}_\nu$ reads:
\be 
\Phi^{LST\Pi}_\nu = \sum^{N_H}_{kl} D_{kl}^\nu(\rho) | kl, LST\Pi \ket ,
\label{adexp}
\ee
with the basis element  given, for $L=0$, by 
\be
| kl, 0S\Pi \ket = \sum_{i=1}^3 \left[ ^{(2)}P^{l,l}_{k}(\Omega_i) \otimes T_i \otimes S_i \right] ,
\label{basis}
\ee
where $S_i$ ($T_i$) indicates the coupling of particles $jki$ to a state of total spin
 $S$ (total isospin $T$),  and the
 hyperspherical polynomial is written as 
(see for instance Ref. \cite{hh2,hh3} for more details):
\be
^{(2)}P^{l,l}_{k}(\Omega) = N_{kl} (1-z^2)^{(l/2)}P_k^{l+1/2,l+1/2}(z)
 P_l(\mu)\;\; ,
\ee
where $P^{\alpha,\beta}_k$ is a Jacobi polynomial, $P_l$ is a Legendre
polynomial and $N_{kl}$ is a normalization factor.
The HH so defined are eigenfunctions of the grand-angular operator, 
\be 
G^2 | kl, 0ST\Pi \ket = K(K+4) | kl, 0ST\Pi \ket, 
\ee
where $K$ is the grand-angular quantum number ($K=2k+2l$).

The unknown coefficients $\{ D_{kl}^\nu \}$ in eq. \refeq{adexp}, and the adiabatic 
potential $\{ U_\nu \}$ are obtained as the eigenvectors and eigenvalues, 
respectively, of the following generalized eigenvalue problem
\be
\sum_{kl}^{N_H} \bra k'l',LST\Pi |\ham_\Omega - U | kl,LST\Pi \ket  D_{kl} = 0 .
\label{hhsys}
\ee

In practical calculations the size $N_H$ of the HH basis set 
is increased until convergence is reached for the desired number $N_A$ of adiabatic 
potentials $\{ U_\nu\}$. However, it is well known that the convergence becomes 
harder to achieve the larger the value of $\rho$. The reason for this behavior is 
connected to the specific properties of the HA basis set at large $\rho$. 
Namely, the lowest adiabatic potentials tend to the binding energies of all possible
two-body subsystems, and the associated HA basis elements to the two-body wavefunctions, 
opportunely normalized. The HH expansion is not optimal for reproducing 
wavefunctions with similar characteristics, which become the more localized the larger $\rho$ . 
This convergence problem can be further enhanced by the
presence of a hard core repulsion in the two-body potential. 
If the calculation is to be limited to the three-body bound states, and in absence of 
very extended ones such as the Efimov states, the limited radius of convergence of 
the HH expansion does not constitute a problem.
When the calculation is extended to the continuum energy region, however, 
the accurate determination of the adiabatic curves and
functions at very large $\rho$ becomes essential for the convergence of the results.
In order to overcome this problem Blume
and co-workers \cite{bge00} advocate the use of splines, which at large $\rho$
converge significantly faster than the HH.
Alternatively,
when $\rho$ is much larger than the range of the two-body interaction, 
approximations for the HA basis elements and potentials
can be obtained by solving a non-homogeneous  
one-dimensional differential equation.
 A brief illustration of this second approach is summarized below,
 based on the work of Nielsen and co-workers \cite{nielsen}.
 Let us start from the definition of the reduced amplitudes $\phi_\nu$
\be
\Phi^{LST\Pi}_\nu = \sum_{i=1,3}^3 \Phi_\nu^{(i)} =
\sum_{i=1}^3 \frac{\phi_\nu(\theta_i,\rho)}{\cos\theta_i\sin\theta_i}.
\ee
each one having the set of quantum numbers ${LST\Pi}$. They
are the solutions of the Faddeev equations,
 that for $s$-wave potentials read
\bea
&& \left(-\frac{\hbar^2}{2 m \rho^2}\frac{d^2}{d\theta^2_i} + 
V(\sqrt{2}\rho\cos\theta_i) 
- \lambda_\nu(\rho) \right) \phi_\nu(\rho,\theta_i) = \nb && 
-{\cos\theta_i\sin\theta_i} V(\sqrt{2}\rho\cos\theta_i) 
\int_{-1}^{1}d\mu_i\left(\Phi_\nu^{(j)}+\Phi_\nu^{(k)}\right )
\label{fadeq}
\eea
where $\lambda_\nu(\rho) = U_\nu(\rho) - 4 \hbar^2/(2 m \rho^2)$. 
Defining $r_0=\sqrt2\rho\cos\theta_0$ the range of the (short-range) 
potential, we observe that,
for large values of $\rho$, the potential $V(\sqrt{2}\rho\cos\theta_i)$ 
can be considered different for zero only for values of $\theta_i$ in the
interval $\theta_0 \le \theta_i \le \pi/2$, which is the smaller the larger
 $\rho$.
Accordingly, the above equation has two regimes depending the values of
$\theta_i$. It is homogeneous for $\theta_i<\theta_0$. For values
in which the potential is not zero we have to evaluate the non-homogeneous 
term which depends on the amplitudes $j,k$.
 From the relation between the different sets of Jacobi coordinates,
 the region of values of $\theta_i$ where $V$ is different from zero 
correspond to the values 
$\theta_j\approx \pi/6$ and $\theta_k\approx \pi/6$. 
 In this region each of these amplitudes is governed by the 
corresponding homogeneous Faddeev equation. For example, for the
$j$-amplitude, the possible solutions depending on the value
of $\lambda_\nu$ are
\bea
\phi_\nu(\rho,\theta_j)=A\sin(k_\nu\theta_j)  &   \lambda_\nu> 0   \cr
\phi_\nu(\rho,\theta_j)=A({\rm e}^{k_\nu\theta_j}-{\rm e}^{-k_\nu\theta_j})
 & \lambda_\nu < 0 \;\;\; ,
\eea
and similarly for the $k$-amplitude, where $k_\nu^2=2 m |\lambda_\nu|/\hbar^2$.
Replacing these expressions in the Faddeev equation \refeq{fadeq},
its asymptotic form can be obtained:
\be
\left(-\frac{\hbar^2}{2 m \rho^2}\frac{d^2}{d\theta^2} + V(\sqrt{2}\rho\cos\theta)
- \lambda_\nu(\rho) \right) \phi_\nu(\rho,\theta) =
V(\sqrt{2}\rho\cos\theta) A f(\rho,\theta)
\label{fedeq}
\ee

When the equation describes a two-body bound state with a third
 particle far away, $\lambda_\nu$ is
negative and tends to the two-body bound state energy. The corresponding
non-homogeneous term is
\be
f(\rho,\theta)= - 2 \frac{e^{k(\pi/2-\theta)}-e^{-k(\pi/2-\theta)}}{k}\frac{e^{k\pi/6}
              -e^{-k\pi/6}}{\sin{(\pi/3)}}.
\ee
 For positive values of $\lambda_\nu $  the
adiabatic functions describe asymptotically three free particles and 
\be
f(\rho,\theta)=  - \frac{8 \sin{(k \pi/6)}}{\sqrt{3}} \sin{[k(\pi/2-\theta)]}/k .
\ee
 $A$ is a normalization constant to 
be determined from the solutions. 
The boundary conditions for the functions $\phi_\nu$ are 
$\phi_\nu(\rho,0)=\phi_\nu(\rho,\pi/2)=0$, which determine completely the
solutions of eq. \refeq{fedeq}.

In practical applications the adiabatic potentials $\{U_\nu\}$ and the HA basis elements
$\{\Phi^{0ST\Pi}_\nu\}$ are obtained as solutions of eq. \refeq{hhsys} for $\rho \le \rho_0$ 
and of eq. \refeq{fedeq} for $\rho> \rho_0$, respectively. The matching point $\rho_0$ needs 
to be chosen larger than the range $r_0$ of the two-body potential $V$. 
There is a zone around the matching point in which, for a sufficient large 
value of $N_H$, the solutions obtained from the HH expansion or
by solving eq. \refeq{fedeq} for each value of $\nu$ become indistinguishable 
from each other. In this way we link the definitions of $\rho_0$ and $N_H$
 as the values for which 
 the solutions of eqs.\refeq{hhsys} and \refeq{fedeq} can be accurately
 matched. In fact, if the functions $\phi_\nu$ obtained by solving eq.
 \refeq{fedeq} are themselves expanded into the HH basis, the coefficients of
 this expansion can be individually matched to the equivalent coefficients 
obtained through solving eq. \refeq{hhsys} for the same value of $\rho$. 

In the following we discuss the solutions of the
the system of coupled differential equations \refeq{usys} in the
case of bound states. The hyperradial functions $\{ u_\nu^n\}$ can be
expanded into normalized generalized Laguerre polynomials times and
exponential function \cite{abr}:
\be
u_\nu^n(\rho)=\sum_{m=0}^{N_p-1} A_{m\nu}^{n} L^{(5)}_m (\beta \rho) \exp{[-\beta \rho/2]}, 
\label{lagbasis}
\ee
where $\beta$ is a non-linear parameter which can be used to improve the
 convergence of the expansion \cite{pb1}. The coefficients $\{A_{m\nu}^{n}\}$
 can be found by means of the Rayleigh-Ritz variational principle,
 whose implementation requires the solution of the following eigenvalue
 problem:
\be
\sum_{m\nu} \bra m'\nu' |\ham - E | m\nu \ket  A_{m\nu} = 0 ,
\label{bsystem}
\ee
where the ortonormalized
basis element $| m \nu \ket$ is defined as
\be
| m \nu \ket =
L^{(5)}_m (\beta \rho) \exp{[-\beta \rho/2]} 
\Phi^{0ST\Pi}_\nu(\rho,\Omega)  .
\ee
The size of the variational problem is 
 $M=N_p \times N_A$, where $N_A$ is the number of adiabatic basis 
functions retained in expansion of eq. \refeq{adbasis}, and $N_p$ is the number of 
Laguerre polynomials used in expansion of eq. \refeq{lagbasis}.
For sake of simplicity all functions $u_\nu^n$ are expanded using 
the same number of Laguerre polynomials, although this is not strictly
 necessary. The eigenvalues 
$\{ E_n^{(M)}\}$ ($n=1,2,\dots$) represent upper bounds to the eigenvalues 
of the Hamiltonian problem \refeq{hhc} and converge towards them
 monotonically as $M$ is increased.
 The associated set of coefficients $\{A_{m\nu}^n\}$ provide 
approximations to the system wavefunctions.

As it has been mentioned before, there is a complete equivalence between 
the two methods if they include the same number of HH functions. 
In fact the expansion for $\Psi$ in eq. \refeq{adbasis} can be written also as:
\be 
\Psi^{LST\Pi}_n = \sum_{kl}^{N_H} w^{n}_{kl}(\rho) |kl,LST \Pi \ket ,
\ee
and from eq. \refeq{adexp} the following relation can be obtained
\be 
 w^n_{kl}(\rho)= \sum_{\nu}^{N_A} u^{n}_\nu(\rho) D^\nu_{kl}(\rho) \,\, .
\ee
If $N_A$ is set equal to $N_H$
 the matrix $D^\nu_{kl}$ represents a unitary transformation 
between the HA and HH basis sets, therefore
 the two expansions must produce identical sets of eigenvalues and eigenvectors.
Consequently, if in a specific problem, the desired accuracy is reached 
 using $N_H$ HH basis functions, 
 the use of a larger number of 
HH basis elements in the expansion of the adiabatic basis functions 
 is superfluous.
 However, we can expect
that the number of adiabatic functions $N_A$ needed to reach the
same accuracy will be $N_A\ll N_H$. This is because the HA functions
have been optimized to the specific Hamiltonian 
 problem by solving eq. \refeq{adeq}
for each value of the hyperradius. We would like to stress the fact that
the equivalence between the HH and the HA method using a tractable
number $N_H$ of HH functions applies in presence of deep bound states.
When shallow bound states, as Efimov states, are present the situation 
changes and a direct application of the HH method encounter the problem 
of the inclusion of a very large number of basis states in the expansion
of the wavefunction. This is related to the
correct description of the adiabatic potentials in the
asymptotic regime. 
In this case the use 
of the asymptotic form of the Faddeev equations given
above proves to be extremely useful, as for example in the
solution of three Helium atoms system~\cite{nielsen}.

\section{Scattering Observable Calculations}
\label{scatt}

In this section we apply the HA expansion to the study of 
continuum states of a three-body system. The case considered will 
be the scattering of one particle colliding other two forming
a dimer, at energies below the three-body breakup threshold.
 The wavefunction for the system 
can be written as
\be
\Psi = \Psi_c + \Psi_a ,
\ee
where the first term is $\lleb^2$ and 
describes the system configurations in which the three particles are all 
close
to each other. The second term represents the solution of the Schroedinger
equation in the asymptotic region in which the incident particle
does not interact with the other two ( the discussion will be limited 
to short range potentials). Moreover,  we will consider the case of a 
two-body interaction that supports only one dimer bound state of energy 
$E^{2b}$. Accordingly, we will consider energies $E^{2b} \le E < 0$.

The explicit form of the term $\Psi_a$ 
depends on the energy $E$ of the system.
However, the particular choice of the function $\Psi_a$ is rather arbitrary,
as it can be modified by adding or subtracting any $\lleb^2$ function. 
In the following we will consider and compare two different expressions
for the asymptotic function $\Psi_a$.
Practical applications will be shown for the case of nucleon-deuteron
scattering using the semi-realistic $s$-wave MT-III potential, 
as the repulsive core of the potential allows a better understanding 
of the numerical problems associated with the method's implementation.

\subsection{Scattering below Break-up: Method 1}

The $\Psi_a$ term must describe the asymptotic state of 
the dimer plus a third particle. Therefore, the most natural choice for 
this term leads to building two independent and 
symmetrized states, that for $L=0$, read as follows:
\be
\Omega^R_{ST} = \sum_i \n \frac{g(r_i)}{r_i} 
\frac{\sin{[ k_y y_i]}}{k_y y_i} P_0(\mu_i)|ST \ket , 
\label{ha1r}
\ee
and 
\be
\Omega^I_{ST} = \sum_i \n \phi_d(r_i)
\frac{\cos{[ k_y y_i]}(1-\exp[-\gamma y_i])}{k_y y_i} P_0(\mu_i)|ST\ket .
\label{ha1i}
\ee
The distance between particle $i$ and particles $j,k$ forming a dimer
is $y_i$, $\phi_d(r)$ is the dimer wavefunction of energy $E^{2b}$, 
$k^2_y =4m(E-|E^{2b}|)/3\hbar^2$ and 
 $\n $ is a normalization factor chosen so that 
\be
\bra \Omega^R_{ST} |\ham-E| \Omega^I_{ST} \ket 
 - \bra \Omega^I_{ST} |\ham-E| \Omega^R_{ST} \ket = 1/2  \,\, .
\label{norma}
\ee
The behavior of the function $\Omega^I_{ST}$ for $y_i\rightarrow 0$ has been
regularized by means of an opportune factor. The constant $\gamma$
can be consider a non linear parameter of the scattering
wave function. The final result should be independent of the value 
chosen for it but a wrong choice can slow down the convergence 
significantly. A reasonable choice could be $\gamma\approx\sqrt{m|E^{2b}|/\hbar^2}$.

A general scattering state is given by defining the following linear combinations
\be
\Omega^0_{ST} = u_{0R} \Omega^R_{ST} + u_{0I} \Omega^I_{ST} , 
\label{omega0}
\ee
and 
\be
\Omega^1_{ST} = u_{1R} \Omega^R_{ST} + u_{1I} \Omega^I_{ST} .
\label{omega1}
\ee
The term $\Psi_a$, having total spin $S$ and total isospin $T$,
can thus be written as 
\be
 \Psi_a = \Omega^0_{ST} + \lscat \Omega^1_{ST}
\label{psia}
\ee
where different choices for the matrix $u$ can be used to define the scattering
matrix $\lscat$ \cite{k1}. Here we will use 
\be
u = \left(
\begin{array}{cc}
i & -1 \\ i & 1
\end{array}
\right)
\ee
defining $\lscat\equiv$ $S$-matrix and $\det u =2 \imath$. Another
possible choice used here corresponds to $u_{0R}=u_{1I}=1$ 
and $u_{1R}=u_{0I}=0$ defining $\lscat\equiv$ ${\cal R}$,
the reactance matrix.
The two representations are related as
\be
{\cal S}= (1+i{\cal R})(1-i{\cal R})^{-1} .
\ee
This identity holds for the exact matrices therefore it can be 
used as a check of the accuracy of the calculation by comparing
the results using both schemes.

At energies below the three-body breakup, the $\Psi_c$ term is $\lleb^2$.
 Accordingly it can be represented by means of an expansion in the same
 $\lleb^2$ basis used for bound states, namely
\be
\Psi_c=\sum_{m\nu} A_{m\nu}|m \nu \ket
\label{lagscat}
\ee
 
From the above definitions we can construct the scattering state as

\be
\Psi = \sum_{m\nu} A_{m\nu}|m \nu \ket + \Omega^0_{ST} + \lscat \Omega^1_{ST}
\ee

The solution of a scattering problem at a given energy requires the
determination of the amplitude $\lscat$ and the linear coefficients
$A_{m\nu}$. To this aim we make use of the
Kohn variational principle \cite{k1} that can be written as
\be
[\lscat] = \lscat - \frac{2}{\det u} \bra \Psi^* | \ham - E | \Psi \ket. 
\ee

The numerical implementation of the variational principle leads to a first
order approximation of the amplitude $\lscat$ obtained through the solution 
of a linear system of equations of size $M+1$, where $M$ is the size of the 
basis set for the expansion of the core part of the wavefunction.
If we define an array of unknowns $(\{A_{m\nu}\}, \lscat )$ of dimension $M+1$,
the linear system can be written as:
\be
\left(
\begin{array}{cc}
 H_{m'\nu',m\nu} &  H_{m'\nu',\Omega^1}  \\ 
 H_{\Omega^1,m\nu} & H_{\Omega^1,\Omega^1} 
\end{array}
\right)
\left(
\begin{array}{c}
A_{m\nu} \\
\lscat
\end{array}
\right)
=
\left(
\begin{array}{c}
 -H_{m'\nu,\Omega^0}  \\
\frac{1}{4}\left( \det u - 2 H_{\Omega^1,\Omega^0} 
- 2 H_{\Omega^0,\Omega^1} \right)
\end{array}
\right) ,
\label{lsystem}
\ee
where $H_{x',x}$ stands for the matrix element
\be
H_{x',x} = \bra x'^* | \ham - E | x \ket .
\ee

The second order estimate for $\lscat$ is then given by
\be
\lscat^{2nd} = \lscat^{1st} - \frac{2}{\det u} \bra {\Psi^{1st}}^* | \ham - E | \Psi^{1st} \ket ,
\label{2ndorder}
\ee
where $\Psi^{1st}$ is the wavefunction obtained 
solving the linear system of eq.\refeq{lsystem}.

Let us now discuss in more detail the structure of eq. \refeq{lsystem}.
The top left part of the coefficient matrix, of dimension $M \times M$
contains the matrix elements used for the bound state calculation when the
scattering state has the same quantum numbers as the bound state (compare it to eq. \refeq{bsystem}).
Otherwise specific states $|m\nu\ket $ having proper quantum numbers
have to be constructed.
The additional matrix elements needing to be computed 
are those between the $\lleb^2$ basis 
functions and the scattering functions,
and among the scattering functions themselves, 
for a total of $2M+4$ different terms. 
The number of such extra terms grows linearly with the basis set size,
and due to the functional form of $\Omega^0_{ST}$ and $\Omega^1_{ST}$, 
they need to be calculated at every different choice of the system 
energy $E$.

The application discussed above employ the HA basis in the
expansion of the $\lleb^2$ $\Psi_c$ term.
Alternatively, $\Psi_c$ could also have been expanded in terms of 
 sole HH functions as 
\be
\Psi_c=\sum_{mkl} A_{mkl}|m kl \ket , 
\ee
where we have defined the ket
\be
| m kl \ket =
L^{(5)}_m (\beta \rho) \exp{[-\beta \rho/2]}\otimes 
|kl, 0ST\Pi \ket .
\ee
After including a sufficient number of Laguerre polynomials, 
both expansions, in terms of HH or HA functions, are equivalent 
leading to the same value
of $\lscat$. Example of this equivalence will be shown and discussed in
the next Section.

\subsection{Scattering below Break-up: Method 2}

An alternative approach considered is represented by a  direct 
solution of eq. \refeq{usys} which represents
a different form of the  three-body Schroedinger equation.
 The bound state solutions have been discussed in Sect.II,
and here we will discuss the scattering solutions below three-body breakup:
 $E^{2b} \le E < 0$. For this purpose 
it is important to determine the boundary conditions to be 
imposed to the functions $\{u_\nu^E(\rho)\}$.  
Firstly, let us observe that at very large $\rho$ the only open channel in 
the system of eqs.\refeq{usys} is the lowest one, and that the system 
uncouples:
\be
\left(-\frac{\hbar^2}{2 m} T_\rho +U_1-E + B_{11} \right)  u_1=0.
\label{usysa}
\ee
At $\rho=0$ corresponds $u_1(0)=0$, whereas
 the boundary conditions at large $\rho$ depend on the specific 
asymptotic forms of the hyperradial potentials $U_1(\rho)$ and 
of the terms $B_{1 1}(\rho)$. A detailed study of their asymptotic 
expressions will be object of a forthcoming publication \cite{nota}. 
For the purpose of this work it suffices to say that 
\be
\left( -\frac{\hbar^2}{2 m} T_\rho +U_1-E + B_{11} \right)  
u_1 \rightarrow \left( \frac{d^2}{d\rho^2} + k_\rho^2 
+ o(\rho^{-3})\right) (\rho^{5/2} u_1),  
\label{asym1}
\ee
where  the wavenumber $k_\rho$
is defined from the relation:
\be
E=E^{2B} + \frac{\hbar^2}{2 m} k_\rho^2.
\ee
The boundary conditions associated with $u_1$ thus are 
\be 
u_1(0)=0 ,  \ \ \ \  \lim_{\rho\rightarrow \infty} u_1(\rho) \rightarrow \tilde{u_1} = \frac{\sin{(k_\rho \rho)}}{\rho^{5/2}} 
    + \tan\delta\frac{\cos{(k_\rho \rho)}}{\rho^{5/2}},
\label{asym}
\ee
all other $u_\nu\rightarrow 0$ sufficiently fast,
 as $\rho\rightarrow\infty$. Furthermore,
the lowest adiabatic function 
$\Phi_1^{0ST\Pi}(\rho,\Omega)\rightarrow \rho^{3/2}\phi_d(r)|ST\ket$ at very
large values of $\rho$~\cite{ffs88}. Therefore, the asymptotic behavior
of the scattering wave function in terms of the adiabatic basis results:

\be
 \Psi= \sum_\nu u_\nu(\rho)\Phi_\nu^{0ST\Pi}(\rho,\Omega) \rightarrow 
% \tilde{u_1} \rho^{3/2}\phi_d(r)|ST\ket=
 \phi_d(r)\left[ \frac{\sin{(k_\rho \rho)}}{\rho}
    + \tan\delta\frac{\cos{(k_\rho \rho)}}{\rho}\right]|ST\ket .
\label{eqas}
\ee
In the limit $\rho \rightarrow \infty$ the relation $k_y y \approx k_\rho \rho$ holds as $r$ is constrained by the finite size of the dimer wavefunction,
 therefore $r/\rho \ll 1$.
Consequently eq. \refeq{eqas} represent the asymptotic limit of
  $\Omega^R_{ST}+\tan \delta \Omega^I_{ST}$,  for $\rho\rightarrow \infty$.
 The full equivalence between the above expression for the asymptotic
wavefunction and that one given by eqs. (\ref{ha1r},\ref{ha1i}) 
can be established by noticing 
that the $\tilde{u_1}$ constitutes the leading term in the expansion of 
 $\Omega^R_{ST}$  and $\Omega^I_{ST}$
in terms of the small parameter $r/\rho$~\cite{fab1}, which yields
\be
\bra \Omega^R_{ST} |\Phi_1 \ket \approx \frac{\sin{[k_\rho \rho]}}{\rho^{5/2}} 
     + {\cal O}(\rho^{-7/2}),
\label{omexp1}
\ee
and
\be
\bra \Omega^R_{ST}|\Phi_\nu \ket \approx \frac{\cos{[k_\rho \rho]}}{\rho^{5}} 
  + {\cal O}(\rho^5) (\nu > 1) .
\label{omexpnu}
\ee
and a similar expansion yields for $\Omega^I_{ST}$. From the above
discussion, we can define an alternative asymptotic term $\Phi_a$
as combination of the following functions:
\be
\Omega^R_{\rho,ST} = \sqrt{\frac{m}{2\hbar^2 k_\rho}} (1-\exp[-\gamma \rho])^{\eta}
\frac{\sin{[k_\rho \rho]}}{\rho^{5/2}} \Phi_1(\Omega,\rho),
\label{ha2r}
\ee
and
\be
\Omega^I_{\rho,ST} = \sqrt{\frac{m}{2\hbar^2 k_\rho}} (1-\exp[-\gamma \rho])^{\eta}
 \frac{\cos{[k_\rho \rho]}}{\rho^{5/2}} \Phi_1(\Omega,\rho),
\label{ha2i}
\ee
where the factor $(1-\exp[-\gamma \rho])^{\eta}$ is introduced
as usual to regularize the behavior of the functions for $\rho\rightarrow 0$ 
(in practical calculations we have set $\eta=4$), 
and the functions are normalized as in eq. \refeq{norma}.
The same approach as in the previous Section can now be applied where.
Accordingly the scattering wave function can be written as
\be
\Psi=\sum_{M\nu}B_{m\nu}|m\nu\ket + \Omega^0_{ST} +{\cal S}
 \Omega^1_{ST}  \;\;\; 
\ee
where the asymptotic part is now given in terms of $\Omega^R_{\rho,ST}$ and  
$\Omega^I_{\rho,ST}$, and 
the core part $\Psi_c$ is expanded onto the HA basis times a set of 
$\lleb^2$ functions $u_\nu(\rho)$.
We can refer to this expansion as HA2. 

This approach is justified as the neglected terms in the $r/\rho$ expansion 
of $\Omega^R$ and  $\Omega^I$ do not carry flux and can be incorporated into 
the unknown term $\Psi_c$.
The approximated expression for the term $\Psi_a$ allows to speed up the 
calculation significantly as there is no need to calculate 
 the overlap integrals between the HA basis functions and the asymptotic 
functions as in eq. \refeq{lsystem}. 
On the other hand,
 its implementation suffers from the following problems.
 At intermediate distance the expansion on $r/\rho$ of the asymptotic 
functions converges very slowly, resulting in  
a  large number of HA functions which need to be taken into account.
At large $\rho$, the implementation of the functions of eqs. 
(\ref{ha1r},\ref{ha1i}) 
 results in a very awkward behavior of the $\lleb^2$ term $\Psi_c$. 
Continuing the expansion of eq. \refeq{omexp1}, for instance, it is possible 
to show that the next term is  $\cos{[k_\rho \rho]}/\rho^{7/2}$, which   
imposes the asymptotic behavior that the function $u_1$ has to reproduce.
This particular functional form is very slow decaying, and it is particularly 
hard to reproduce with a polynomial expansion.
This problem is further enhanced by the presence of oscillations associated
with cosine and sine terms.

In order to solve the linear system \refeq{lsystem} taking into account
the oscillatory behavior of the hyperradial functions for large
$\rho$ values, we have implemented
a Discrete Variable Representation (DVR) scheme \cite{dvr}
rather than the standard variational approach.
In a previous work \cite{lbk04} we have shown how to combine the variational 
Kohn principle with a DVR scheme, for the case of a two-body system, 
which corresponds to a single one-dimensional differential equation.
In this work we have a set of N$_A$ one-dimensional coupled differential 
equations. Therefore we define a $(N_A M +1)\times (N_A M +1)$
 unitary transformation
 matrix $\mdvr$ which is a direct product of $N_A+1$ matrices
\be
\mdvr = \mdvr^{1d} \otimes \mdvr^{1d} \otimes \dots \otimes 1 , 
\ee
where $\mdvr^{1d}$ is a $M\times M$ unitary matrix associated 
to a customary one-dimensional DVR of size $N_{DVR}=M$ built in $\rho$:
\be
\mdvr^{1d}_{ij} =  L^{(5)}_{i} (t_j) \exp{[-t_j/2]} \sqrt{w_j} ,
\ee
 where $t_j$ and $w_j$ are the appropriate quadrature points and weights.
By mean of a parameter $\beta$, the end quadrature point $t_{N_{DVR}}$ can be 
associated to different physical 
values $\rho_{max}$, by setting $t_j = \beta \rho_j$.
In this fashion we can constrain the quadrature points to be distributed
 between $0$ and $\rho_{max}$. 

\section{Numerical Applications}

In order to illustrate the method outlined in the previous Sections we
present two applications to the $n-d$ system in a quartet state ($S=3/2$).
The potential energy of the system is taken as the sum of three pairwise
potentials. We consider the MT-III interaction $V_{MT-III}$
for which benchmarks results exist in the literature~\cite{cpf89}.
It reads:
\be
V_{MT-III}(r) =  \left( 1438.72 \exp{[-3.11 \, r]} - 626.885 \exp{[-1.55 \, r]}\right)/r.
\ee
To make contact with the results of Ref.~\cite{fab1},
we have also used the Gaussian potential (named $V_G$):
\be
V_G(r) = -66.327 \exp{[-(0.64041 \, r)^2]},
\ee
For both potentials we assume nuclear distances in fm and energies in MeV.
The nucleon mass used is such that $\hbar^2/m=41.47$ MeV fm$^2$. Furthermore,
we consider both potentials as acting only on the $l=0$ two-body partial wave.

The potential $V_G$ supports one deuteron bound state, with zero angular
momentum, of energy $E_{2b}=-2.22448$ MeV. The zero-energy 
scattering length is $a_s=5.4208$ fm, whereas for the MT-III potential the 
values are $E_{2b}=-2.23069$ MeV, and $a_s=5.5132$ fm.

For the potential $V_G$ we consider the three-body system with quantum 
numbers $\Pi=a$, $T=1/2$ and $S=1/2$, whereas for $V_{MT-III}$ 
$\Pi=a$, $T=1/2$ and $S=3/2$. As the potentials are projectors on $s-$wave, 
 the index $l$ in eq. \refeq{basis} is restricted to the value $l=0$, 
and the index $k$ can take the values $k=0,2,3,4,5,\dots,\infty$ in the first 
case and $k=1,2,3,4,5,\dots,\infty$ in the second case. 

\subsection{Bound states}

\begin{figure}[htb]
\begin{tabular}{cc}
\includegraphics[scale=0.20,angle=-90]{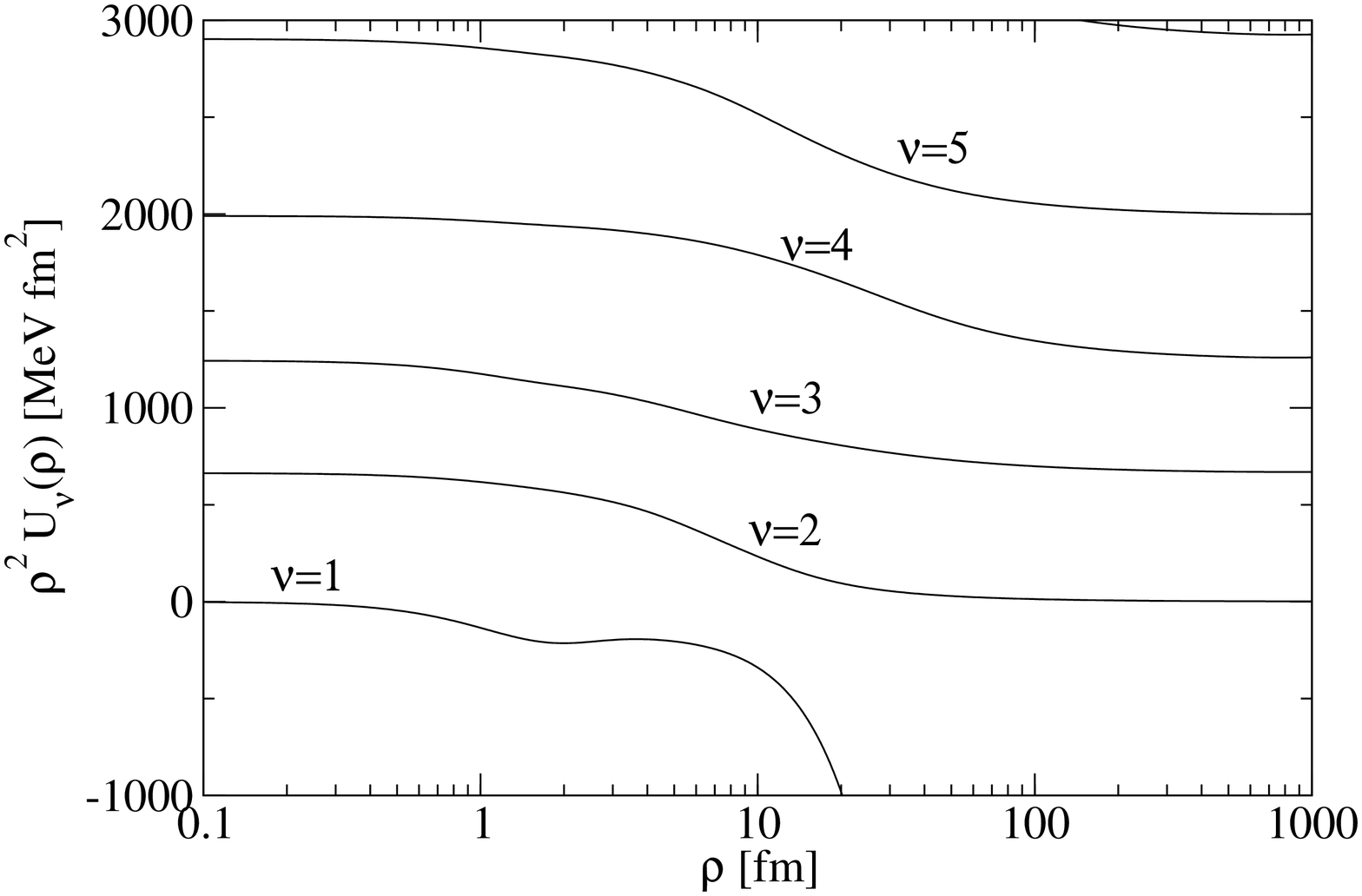} &
\includegraphics[scale=0.20,angle=-90]{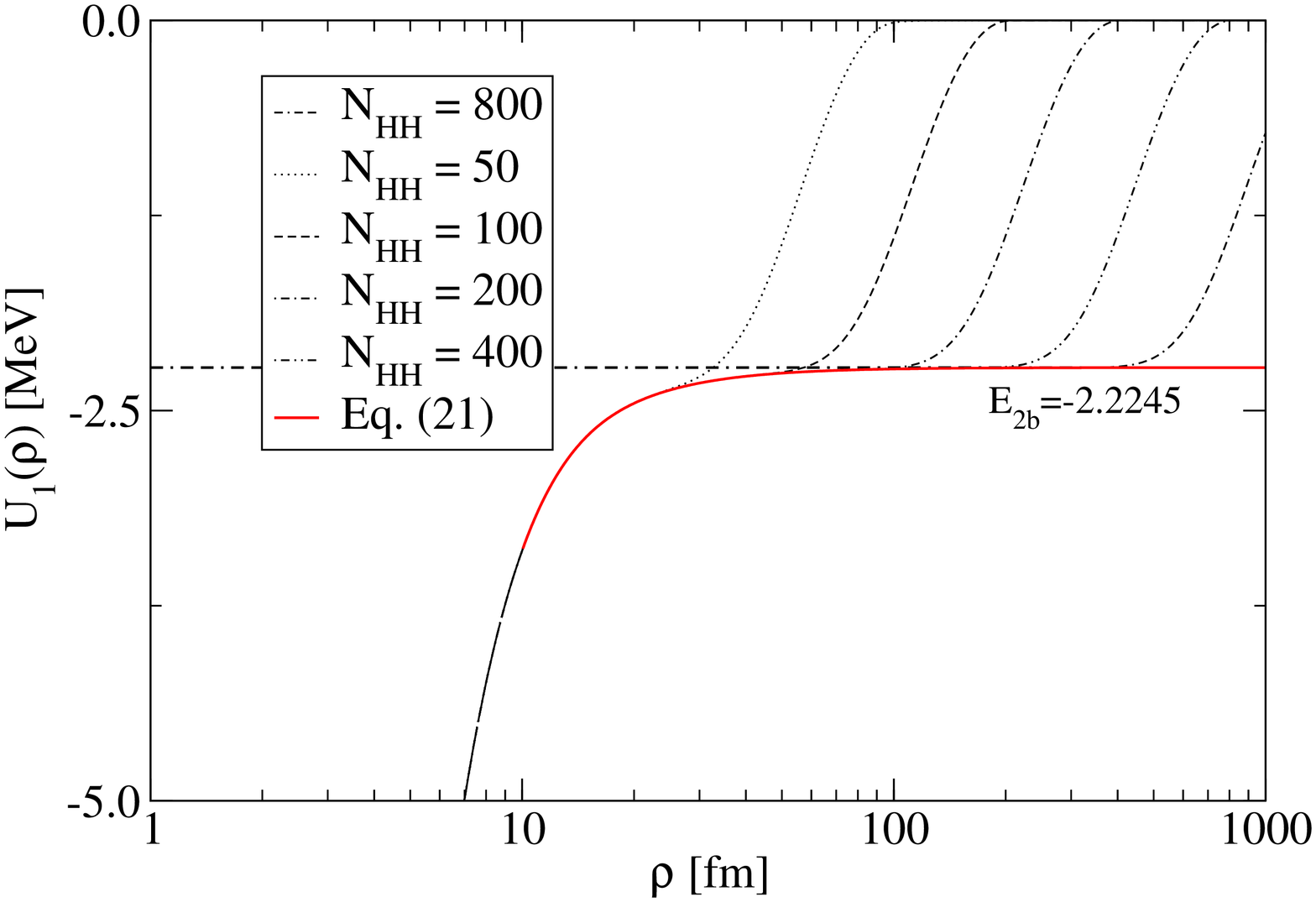} 
\end{tabular}
\caption{The top panel shows the lowest adiabatic curves $U_\nu(\rho)$ for different values of $\rho$.
 In order to display the behavior at large $\rho$ the curves are multiplied by a factor $\rho^2$.
 The lowest curve thus tends to $E_{2b}\rho^2$, and the others to the spectrum $\hbar^2 K(K+4)/m$, 
with $K=0,2,3,4,\dots$ for $\nu=2,3,4,5,\dots$. The bottom panel shows the convergence of the lowest 
adiabatic curve $U_1(\rho)$ as a function of the number of HH used in the expansion of
 eq. \refeq{adexp}. The asymptote at $E_{2b}=-2.2245$ MeV is plotted for comparison.}
\label{fig1}
\end{figure}

Figure \ref{fig1} shows, in the upper panel, the lowest adiabatic curves 
$U_\nu(\rho)$ calculated for the $V_G$ potential. In order to highlight 
their asymptotic behavior,
the curves have been multiplied by a factor $\rho^2$.
The lowest curve $U_1(\rho)$  thus tends to the deuteron energy times $\rho^2$,
whereas the upper curves tend to the free HH spectrum, that is 
$4k(k+2)\hbar^2/m$ with $k=0,2,3,4,5,\dots $ for $\nu=2,3,4,5,6,\dots$. 
The value  $k=1$
is not allowed as there is no completely symmetric HH with $k=1$ and $l=0$. 
Subsequently, the adiabatic function $\Phi_1(\rho,\Omega)$ tends 
to the deuteron wavefunction, whereas $\Phi_\nu$, $\nu > 1$, to the HH
functions, with the appropriate normalization factors.
The lower panel shows the convergence of the lowest curve $U_1(\rho)$ as a 
function of the number of HHs employed in the expansion of eq. \refeq{adexp}.
It shows that the larger $\rho$ becomes, the larger the expansion basis must 
be in order to properly describe the function $\Phi_1$. In practice, the radius
of convergence of expansion \refeq{adexp} increases rather slowly when the 
basis set size is increased.
The reason for this behavior is that when $\rho$ is increased
the function $\Phi_1$ becomes more and more localized in the 
hyperangular phase-space, therefore its 
description by means of the HH requires a larger and larger basis set size.
This behavior is not connected with any particular feature of the potential
used in this specific calculation but it can be considered a general one, 
as it is induced by the geometric localization of the deuteron wavefunction 
in connection with the HH expansion. The thick curve is the solution of
eq.~\refeq{fedeq} starting at $\rho=20$ fm. For large values of $\rho$,
the corresponding eigenvalue reproduces the two-body binding energy
$E_{2b}$.

The description of a three-nucleon bound state using a central potential
has to been taken as a homework problem and preliminary to check the
usefulness of the HA basis to treat scattering states, in comparison 
to the HH expansion. 
The $V_G$ potential predicts two bound states in the three-body system,
a very deep ground state and a very shallow excited state. Table \ref{tab0} 
reports the convergence patterns 
for the upper bounds $E_1^{N}$ and $E_2^{N}$ to 
the two bound states supported by the  potential $V_G$, as a function of the 
number $N$ of HA and HH basis elements. 
The HA functions were expanded in 80 HH functions which is the
number required for the HH expansion to describe accurately the deep and shallow bound 
states. The number of HH functions necessary to obtain a full convergence of 
the energy for the deep bound state is much smaller, around 10 functions.
The most striking feature to be observed in the table is the much rapid 
convergence of the HA basis expansion compared to the HH. Not only full 
convergence can be achieved with a basis which is one order of magnitude
smaller, 
but already the inclusion of only one basis element yields an energy for 
the excited state within 90$\% $ of its converged value.

\begin{table}[hbt]
%%\beforetab

\begin{tabular}{lllclll} \hline \hline 
  & \multicolumn{2}{c}{n=1} & \hspace{2cm}& & \multicolumn{2}{c}{n=2} \\ \cline{2-3} \cline{6-7}
$N$ & HH & HA  & & $N$ & HH & HA  \\ 
1  & -21.5808 & -22.0520 & &  1 &   0.0620 &  -2.3484  \\
2  & -21.9567 & -22.0850 & &  4 &  -0.9576 &  -2.3627  \\ 
3  & -22.0694 & -22.0873 & & 10 &  -2.0348 &  -2.3632  \\  
4  & -22.0805 & -22.0874 & & 20 &  -2.3036 &  -2.3632  \\  
5  & -22.0852 & -22.0874 & & 30 &  -2.3474 &  -2.3632  \\  
6  & -22.0869 & -22.0874 & & 40 &  -2.3582 &  -2.3632  \\  
7  & -22.0872 & -22.0874 & & 50 &  -2.3615 &  -2.3632  \\  
8  & -22.0873 & -22.0874 & & 60 &  -2.3626 &  -2.3632  \\  
9  & -22.0874 & -22.0874 & & 70 &  -2.3631 &  -2.3632  \\  
10 & -22.0874 & -22.0874 & & 80 &  -2.3632 &  -2.3632  \\
\hline \hline
%\hline \hline
\end{tabular}
%\aftertab
\caption{Patterns of convergence for the three-nucleon bound states obtained with the $V_G$ potential, as a function of the number $N$ of hyperangular basis functions included in the expansion. The HA basis elements were calculated with 80 HH, $\beta=1.6$ fm$^{-1}$, and 33 Laguerre polynomials were employed in the expansion of eq. \refeq{lagbasis}. Note the different scales for the ground and excited state patterns of convergence.}
\label{tab0}
\end{table}

\subsection{Scattering States}

In the following, results obtained combining the HA basis expansion with the 
expressions of eqs.(\ref{ha1r},\ref{ha1i}) are given and will be referred to as HA1.
Table \ref{tab1} reports the full patterns of 
convergence of the $L=0,S=3/2$ MT-III phase shift 
$\delta$, at $E_{cm}=1$ MeV, as a function of the number of Laguerre polynomials 
$N_p$ used in expanding the hyperradial functions
in eq. \refeq{lagscat} and the number $N_A$ of adiabatic channels included. 
The HA functions have been calculated using 200 HH functions. This number
of HH functions is sufficient to accurately describe the phase shifts below
the three-body breakup.
From the table it can be seen that the convergence requires a rather high number
of HA basis elements, more than 100, 
whereas $12$ Laguerre polynomials are enough to achieve final convergence.

\begin{table}[htb]
%\beforetab

\begin{tabular}{llllllll} \hline \hline
$N_p \backslash N_A$     &   20     & 40      & 60      & 80      & 120     & 160     & 200     \\
 \hline
5    &  -55.974 & -55.912 & -55.902 & -55.898 & -55.897 & -55.896 & -55.896 \\
9    &  -55.937 & -55.879 & -55.870 & -55.867 & -55.865 & -55.864 & -55.864 \\ 
13   &  -55.932 & -55.878 & -55.868 & -55.865 & -55.864 & -55.863 & -55.863 \\
17   &  -55.934 & -55.878 & -55.868 & -55.865 & -55.863 & -55.863 & -55.863 \\ 
21   &  -55.932 & -55.878 & -55.868 & -55.865 & -55.864 & -55.863 & -55.863 \\
25   &  -55.933 & -55.878 & -55.868 & -55.865 & -55.864 & -55.863 & -55.863 \\
29   &  -55.932 & -55.878 & -55.868 & -55.865 & -55.864 & -55.863 & -55.863 \\
33   &  -55.931 & -55.878 & -55.868 & -55.865 & -55.864 & -55.863 & -55.863 \\ 
\hline \hline
\end{tabular}
%\aftertab
\caption{Convergence of the phase-shift $\delta$ in function of the number of Laguerre polynomials $N_p$ (see eq. \refeq{lagbasis}) and of the size $N_A$ of the HA basis set, at an incident energy of $E=1.00$ MeV. The HA basis is calculated with 200 HH elements. The non-linear parameter was fixed to $\beta=1.9$ fm$^{-1}$.}
\label{tab1}
\end{table}

In order to analyze deeply the pattern of convergence, in
Table \ref{tab2} results obtained by means of the HH expansion
 \cite{pap1} are compared
to those obtained with the HA approach. 
In each row of the table $N_A$ indicates the
number of HH functions used in the calculation and the number of HA functions
used calculated using 200 HH functions. As already pointed out, for the special case of 
$N_H=N_A$ the two expansions are equivalent and the results become identical, provided 
that a sufficiently high number of Laguerre polynomials is employed to describe the 
$\{u_\nu(\rho)\}$ set of functions. Therefore the equivalence can be seen in the last 
row of the table in correspondence with $N_A=200$ (in some cases the
equivalence is reached already at $N_A=160$).
For the case of $E=2.00$ MeV, two patterns of convergence are shown for two different 
HA bases, obtained with 120 HH and 200 HH, respectively. Here the equivalence 
can be seen also at $N_A=120$. In this energy range there is little 
difference in the results obtained with the two bases, for example when 20 or 40 HA basis 
elements are employed. 
To be noticed that the results shown in Table \ref{tab2}
present a different pattern of convergence with respect to the ones given in Table 2 of
Ref \cite{pap1}: the reason is that in the previous paper the $S-$matrix representation
was chosen for the matrix $u$, whereas in this work the $R-$matrix was preferred.
The two choices are equivalent and lead, once convergence is achieved, to the same results. 
We can conclude that although there is some improvement, the table shows that the 
convergence is not speed up significantly by transforming the HH basis into the HA basis. 
This  suggests that the HA basis does not provide as an optimized 
basis for the scattering problem as it does for the bound state problem.

\begin{table}[h]
%\beforetab

\begin{tabular}{c|ccccccccc}  \hline \hline
& \multicolumn{2}{c}{0.20 MeV} &\hspace{2cm}& \multicolumn{2}{c}{1.00 MeV} &\hspace{2cm}& \multicolumn{3}{c}{2.00 MeV} \\ \hline 
$N_A$ & HH & HA1 && HH & HA1 && HH & \multicolumn{2}{c}{HA1} \\ \hline
      &          &         &&         &         &&         &   120   &    200  \\ \hline
20    & -28.263  & -28.312 && -56.913 & -55.931 && -70.741 & -71.594 & -71.597 \\
40    & -28.201  & -28.299 && -55.948 & -55.878 && -71.701 & -71.501 & -71.500 \\
60    & -28.306  & -28.295 && -55.922 & -55.868 && -71.508 & -71.485 & -71.483 \\
80    & -28.296  & -28.294 && -55.872 & -55.865 && -71.483 & -71.480 & -71.478 \\
120   & -28.294  & -28.294 && -55.865 & -55.864 && -71.476 & -71.476 & -71.475 \\
160   & -28.294  & -28.294 && -55.863 & -55.863 && -71.474 &    -    & -71.474 \\
200   & -28.294  & -28.294 && -55.863 & -55.863 && -71.474 &    -    & -71.474 \\ \hline \hline
\end{tabular}
%\aftertab
\caption{ Convergence of the phase-shift $\delta$
 at three different energies below break-up threshold
 for the MT-III potential, in function of the size
$N$  of the basis.
The patterns of convergence for the HH and HA1 methods are shown 
for comparison. The HA basis was calculated employing 200 HH 
 basis elements. For $E=2.00$ MeV the calculation with 120 HH basis elements 
is also shown.
 All calculations employed 33 Laguerre polynomials (see Table \ref{tab1}),
 and $\beta=1.9$ fm$^{-1}$.}
\label{tab2}
\end{table}

Table \ref{tab3a} shows the convergence pattern for the phase-shift 
at E=1.00 MeV, obtained using the HA2 expansion for the asymptotic term.
As anticipated in the previous Section, in order to 
obtain stability in the phase shift, 
we have employed a much larger and finer hyperradial grid, consisting 
of 4153 points, distributed up to $\rho=2000 $ fm. 
At the same time the HA basis set and associated eigenvalues were obtained with a bigger 
number, up to 2000, of HH basis functions, 
or by solving the asymptotic differential
eq.\refeq{fedeq} for $\rho\ge\rho_0$ ($\rho_0=40$ fm). 
This calculation has been performed using the Laguerre polynomials as
 an expansion basis for the hyperradial functions. As anticipated, 
the polynomials are not an appropriate choice to 
reproduce the long range oscillatory behavior of the hyperradial functions.
This can be seen from the poor convergence pattern in terms of $N_p$ as
the number of HA functions increases. For $N_A>8$ more than 100 polynomials are
necessary. Furthermore, the convergence pattern is also poor relative 
to the increase of the number of HA basis elements. Differences with results of
Table \ref{tab1} are remarkable. \\

\begin{table}[hbt]
%\beforetab

\begin{tabular}{r|ccccccc} \hline \hline
$N_p \backslash N_A$     &   4    & 8      & 16      & 24     & 32  & 36    & 40     \\
 \hline
21    &  -57.753 & -57.231 & -57.063 & -57.037 & -57.230 & -57.028 & -57.027 \\
41    &  -57.638 & -56.915 & -56.581 & -56.511 & -56.489 & -56.484 & -56.480 \\ 
61    &  -57.628 & -56.868 & -56.456 & -56.348 & -56.310 & -56.300 & -56.293 \\
81    &  -57.627 & -56.858 & -56.414 & -56.281 & -56.228 & -56.214 & -56.204 \\ 
101   &  -57.626 & -56.855 & -56.399 & -56.251 & -56.188 & -56.169 & -56.156 \\
121   &  -57.626 & -56.853 & -56.393 & -56.237 & -56.166 & -56.145 & -56.129 \\
\hline \hline
\end{tabular}
%\aftertab
\caption{Convergence of the phase-shift $\delta$, using the HA2 method, 
in function of the number
of Laguerre polynomials $N_p$ (see eq. \refeq{lagbasis}) and of the size $N_A$
of the HA basis set, at an incident energy of $E=1.00$ MeV. The HA basis is 
calculated with 2000 HH elements. The non-linear parameter was fixed to
 $\beta=1.9$ fm$^{-1}$.}
\label{tab3a}
\end{table}

\begin{figure}[hbt]
\begin{tabular}{cc}
\includegraphics[scale=0.20,angle=-90]{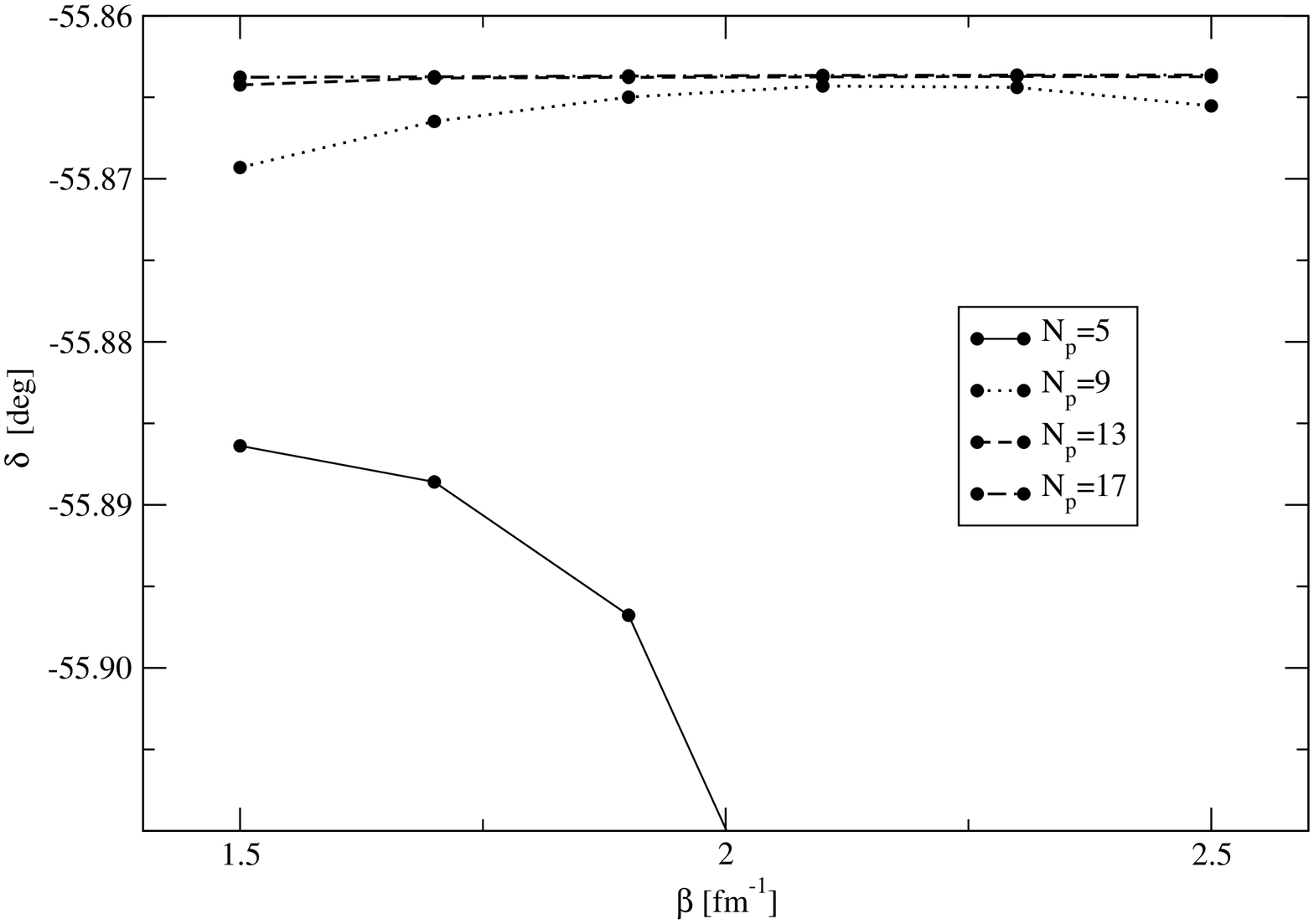}  &
\includegraphics[scale=0.20,angle=-90]{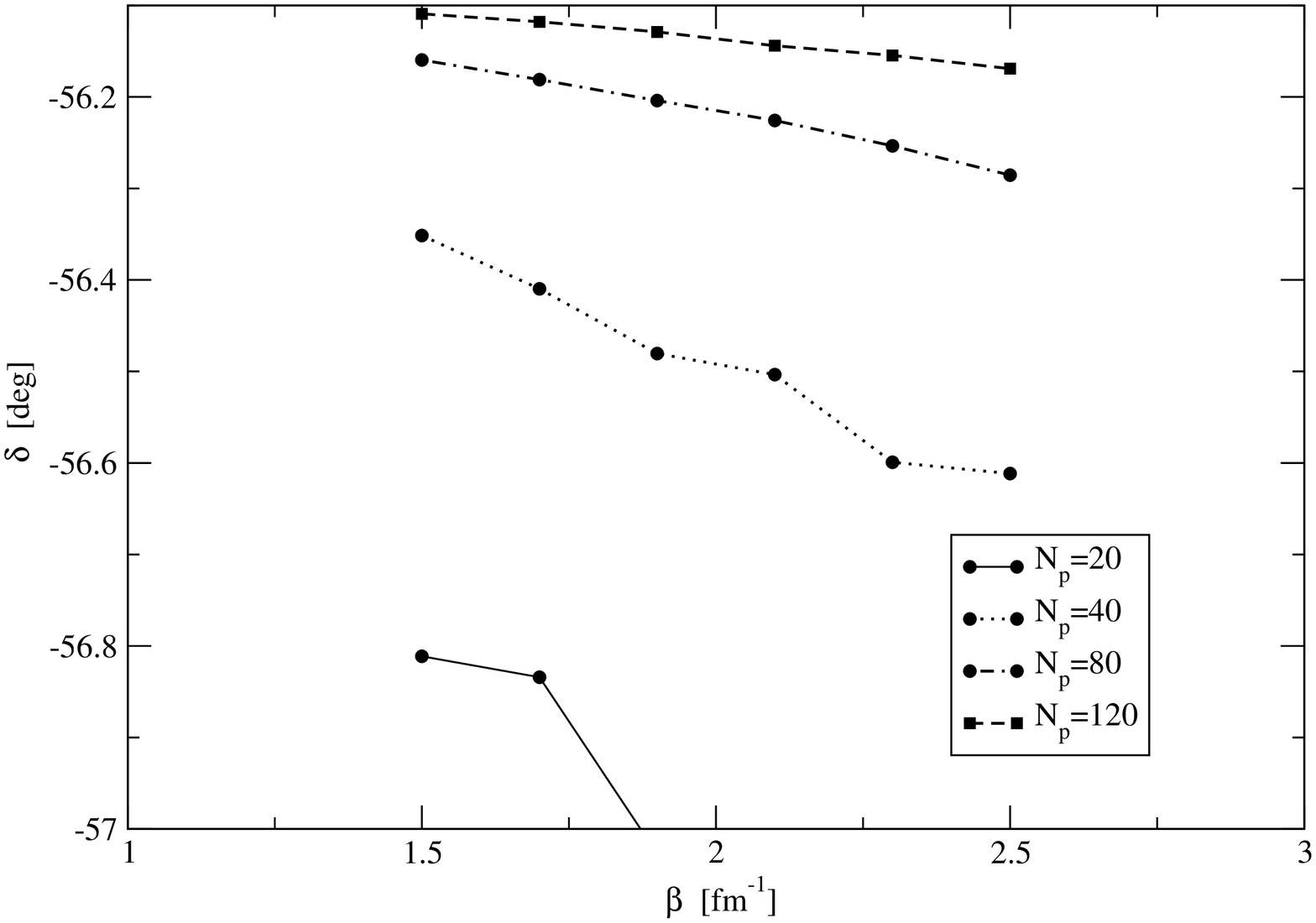}  
\end{tabular}
\caption{The phase-shift $\delta$ 
in terms of different choices of the non-linear parameter $\beta$ and of the 
size of the expansion in Laguerre polynomials.
 The top panel shows the convergence for expansion HA1,
 and the bottom panel for expansion HA2. Note the different scales on the 
$y-$axis of the two graphs.}
\label{fig2}
\end{figure}

Figure \ref{fig2} shows the effect on the 
phase-shift of varying the non-linear parameter $\beta$. The upper panel 
shows results for the HA1 expansion, whereas the lower panel refers to the 
HA2 expansion. Different sizes of the Laguerre basis are shown. In principle, 
for a complete basis set, that is $N_p=\infty$, there should be no effect in 
varying the parameter $\beta$. When the basis set is finite, the stability 
of the result, in this case the phase-shift, with respects to changes of
 $\beta$ is a measure of the completeness of the expansion.
In particular, by comparing the upper and lower panels, one can see that 
the HA1 polynomial expansion of the functions $u_\nu(\rho)$  is 
much more effective than for the case HA2 (also note the different scales of
 the $y-$axis). In the first case, the expansion with $17$ polynomials is 
completely unaffected by changes in $\beta$, whereas in the second case 
even a basis set as large as $120$ polynomials yields significantly 
different results with different choices of $\beta$, indicating that 
the result is far from convergence. 

\begin{table}
%\beforetab

\begin{tabular}{l|cccccccc} \hline \hline
$\rho_{\rm max} \backslash N_{DVR} $ & 100 & 150 & 200 & 250 & 300 && \multicolumn{2}{c}{350} \\ \cline{8-9} 
 & & & & & && 1$^{st}$& 2$^{nd}$ \\ 
%  100 &  -56.376 &  -56.378  &  -56.381  &  -56.383 &  -56.384 &&  -56.386 & -56.385 \\ 
  200 &  -56.179 &  -56.161  &  -56.159  &  -56.159 &  -56.159 &&  -56.161 & -56.160 \\
%  300 &  -56.143 &  -56.115  &  -56.110  &  -56.108 &  -56.108 &&  -56.120 & -56.108 \\
  400 &  -56.124 &  -56.100  &  -56.095  &  -56.093 &  -56.092 &&  -56.091 & -56.092 \\
%  500 &  -56.131 &  -56.091  &  -56.088  &  -56.087 &  -56.086 &&  -56.086 & -56.086 \\
  600 &  -56.096 &  -56.089  &  -56.085  &  -56.084 &  -56.084 &&  -56.085 & -56.083 \\
%  700 &  -56.104 &  -56.087  &  -56.085  &  -56.083 &  -56.082 &&  -56.082 & -56.082 \\
  800 &  -56.119 &  -56.089  &  -56.084  &  -56.083 &  -56.082 &&  -56.080 & -56.081 \\
%  900 &  -56.097 &  -56.085  &  -56.083  &  -56.082 &  -56.082 &&  -56.083 & -56.081 \\
 1000 &  -56.162 &  -56.087  &  -56.082  &  -56.081 &  -56.081 &&  -56.082 & -56.081 \\
% 1100 &  -56.164 &  -56.087  &  -56.083  &  -56.081 &  -56.081 &&  -56.080 & -56.081 \\
 1200 &  -56.149 &  -56.088  &  -56.082  &  -56.081 &  -56.081 &&  -56.082 & -56.081 \\
% 1300 &  -56.128 &  -56.087  &  -56.082  &  -56.081 &  -56.081 &&  -56.079 & -56.081 \\
 1400 &  -56.106 &  -56.084  &  -56.082  &  -56.081 &  -56.081 &&  -56.077 & -56.080 \\
% 1500 &  -56.108 &  -56.088  &  -56.082  &  -56.081 &  -56.081 &&  -56.079 & -56.080 \\
 1600 &  -56.154 &  -56.090  &  -56.082  &  -56.081 &  -56.081 &&  -56.082 & -56.080 \\
%%  100 &   -72.203 &  -72.197 &  -72.196 &  -72.194 & -72.194 &&   -72.191 &  -72.193 \\
%  200 &   -71.911 &  -71.898 &  -71.895 &  -71.893 & -71.892 &&   -71.889 &  -71.892 \\
%%  300 &   -71.959 &  -71.856 &  -71.827 &  -71.823 & -71.821 &&   -71.817 &  -71.821 \\
%  400 &   -71.966 &  -71.828 &  -71.826 &  -71.806 & -71.800 &&   -71.794 &  -71.798 \\
%%  500 &   -71.866 &  -71.856 &  -71.805 &  -71.801 & -71.796 &&   -71.785 &  -71.791 \\
%  600 &   -71.879 &  -71.827 &  -71.810 &  -71.794 & -71.790 &&   -71.780 &  -71.788 \\
%%  700 &   -71.895 &  -71.816 &  -71.801 &  -71.792 & -71.787 &&   -71.779 &  -71.786 \\
%  800 &   -72.002 &  -71.818 &  -71.796 &  -71.788 & -71.785 &&   -71.780 &  -71.784 \\
%%  900 &   -71.901 &  -71.817 &  -71.790 &  -71.788 & -71.784 &&   -71.777 &  -71.782 \\
% 1000 &   -71.961 &  -71.819 &  -71.792 &  -71.784 & -71.784 &&   -71.790 &  -71.782 \\
%% 1100 &   -71.929 &  -71.812 &  -71.794 &  -71.785 & -71.783 &&   -71.779 &  -71.782 \\
% 1200 &   -72.129 &  -71.817 &  -71.793 &  -71.787 & -71.783 &&   -71.774 &  -71.781 \\
%% 1300 &   -72.239 &  -71.814 &  -71.796 &  -71.785 & -71.783 &&   -71.789 &  -71.782 \\
% 1400 &   -72.203 &  -71.828 &  -71.797 &  -71.785 & -71.783 &&   -71.777 &  -71.782 \\
%% 1500 &   -72.104 &  -71.833 &  -71.793 &  -71.786 & -71.783 &&   -71.780 &  -71.782 \\
% 1600 &   -72.388 &  -71.830 &  -71.795 &  -71.785 & -71.782 &&   -71.778 &  -71.781 \\
%% 1700 &   -72.420 &  -71.859 &  -71.797 &  -71.786 & -71.782 &&   -71.780 &  -71.781 \\
% 1800 &   -72.377 &  -71.832 &  -71.793 &  -71.785 & -71.782 &&   -71.774 &  -71.781 \\
 \hline \hline 
\end{tabular}
%\aftertab
\caption{Convergence of the phase-shift $\delta$ at $E=1.00$ MeV, using the HA2 method,
for the MT-III potential, as a function of the number $N_{DVR}$
 of DVR points employed,
 and of the last grid point $\rho_{\rm max}$. Convergence is shown for the
 second order estimate of $\delta$ for all values of M, but the last, where
 both first and second order are shown.} 
\label{tab3}
\end{table}

In order to circumvent this problem we use the DVR technique in the hyperradius
variable.
Table \ref{tab3} shows the convergence, in terms of different
choices of $\rho_{max}$ and the number of DVR points employed, of a case calculation, 
with 40 adiabatic functions, for the MT-III potential and $E=1.00$ MeV.
For the biggest case ($N_{DVR}=350$), we show both the first and second order values of the 
phase-shift obtained by using the Kohn Variational Principle.
In order to obtain a good convergence of the second order value it 
is important that the integral in eq. \refeq{2ndorder} is calculated with a 
very high numerical accuracy. 
The hyperradial grid used in the calculation consists in more than 4000 grid
points up to $\rho=2000$ fm. The use of the DVR technique allowed for stable
results in terms of the hyperradial expansion. The use of 350 DVR points
is equivalent to a calculation with 350 Laguerre polynomials which in general
is much more involved to be carried. However the number $N_A=40$ of HA functions used
in this calculation is not enough to well describe the phase shift. At E=1.00 MeV
the HA1 method as well as the HH method predict $\delta=-55.863$ degrees to be 
compared to the result of the HA2 method, $\delta=-56.081$ degrees, using $N_A=40$.
In order to have an stable result for $\delta$ using the HA2 method, the value
$N_A=120$ has to be considered and $N_{DVR}>350$ since the number of DVR points
has to be increased as $N_A$ increases. The dimension of the HA2 problem is
$N_A\times N_{DVR}$ and is clear that very soon the problem becomes computationally 
unsustainable, unless exceptional computational resources are considered.

\begin{table}[h]
%\beforetab
\begin{tabular}{c|llllllll} \hline \hline
 & \multicolumn{2}{c}{0.20 MeV} && \multicolumn{2}{c}{1.00 MeV} && \multicolumn{2}{c}{2.00 MeV} \\ \cline{2-3} \cline{5-6} \cline{8-9}
 $N_A$  &      HA1  &   HA2  &&    HA1  &    HA2  &&   HA1   &   HA2     \\
  4 &  -28.364 & -29.065 && -56.136 & -57.625 && -72.344  & -71.988 \\
  8 &  -28.340 & -28.739 && -56.038 & -56.852 && -71.965  & -71.871 \\
 12 &  -28.328 & -28.604 && -55.984 & -56.545 && -71.770  & -71.437 \\
 16 &  -28.319 & -28.532 && -55.947 & -56.385 && -71.660  & -71.210 \\
 20 &  -28.312 & -28.487 && -55.922 & -56.286 && -71.597  & -71.070 \\
 24 &  -28.308 & -28.456 && -55.906 & -56.218 && -71.558  & -71.975 \\
 28 &  -28.304 & -28.434 && -55.895 & -56.169 && -71.534  & -71.907 \\
 32 &  -28.302 & -28.417 && -55.888 & -56.133 && -71.518  & -71.855 \\
 36 &  -28.300 & -28.404 && -55.882 & -56.104 && -71.507  & -71.815 \\
 40 &  -28.299 & -28.394 && -55.877 & -56.081 && -71.500  & -71.783 \\ \cline{2-3} \cline{5-6} \cline{8-9}
Table \ref{tab2} & \multicolumn{2}{c}{-28.294} && \multicolumn{2}{c}{-55.863} 
    && \multicolumn{2}{c}{-71.474} \\ \hline \hline
\end{tabular}
%\aftertab
\caption{Patterns of convergence for the two different choice of the asymptotic term, 
in terms of the number $N_A$ of HA basis elements, at three different energies. 
The MT-III potential has been used. 
The last row reports the converged values from Table \ref{tab2}. 
The columns refers to a choice of $\beta=1.9$ fm$^{-1}$ for HA1, and $\rho_{max}=1200$ fm 
for the HA2 expansion. Moreover, the HA1 values are associated to a calculation 
with 200 HH, whereas the HA2 to a calculation with 2000 HH. 
The HA2 results have been obtained with the DVR scheme.}
\label{tab4}
\end{table}

Table \ref{tab4} compares the convergence patterns for $\delta$ at three different energies
 for the two suggested choices for the asymptotic term $\Psi_a$, namely the 
one in eqs. (\ref{ha1r},\ref{ha1i}),
 referred to as HA1, and the one in eqs. (\ref{ha2r},\ref{ha2i}), 
referred to as HA2, in terms of the number $N_A$ of adiabatic channels.
Due to the very large basis sets required to obtain convergence with the HA2 term, 
the pattern of convergence is limited to few channels, less than required to obtain 
a full convergence. The last row reports the converged values from Table \ref{tab2}.
It is possible to see that the expansion HA1 converges faster towards the final 
number, whereas expansion HA2 moves rather slowly. The reason is the difference
in the treatment of the asymptotic wavefunction. In the HA1 method, as well as in the
HH method, the asymptotic configuration described by $\Psi_a$ is reached at 
intermediate distances.  Conversely, in the HA2 the the configuration
described by $\Psi_a$ is reached at much larger values of $\rho$. Furthermore, 
 at intermediate distances, in order to reproduce the correct behavior
a big number of HA functions are needed.

The following figures present important characteristics of the hyperradial 
functions used in the expansion HA2.\\
\begin{figure}[htb]
\begin{tabular}{cc}
\includegraphics[scale=0.20,angle=-90]{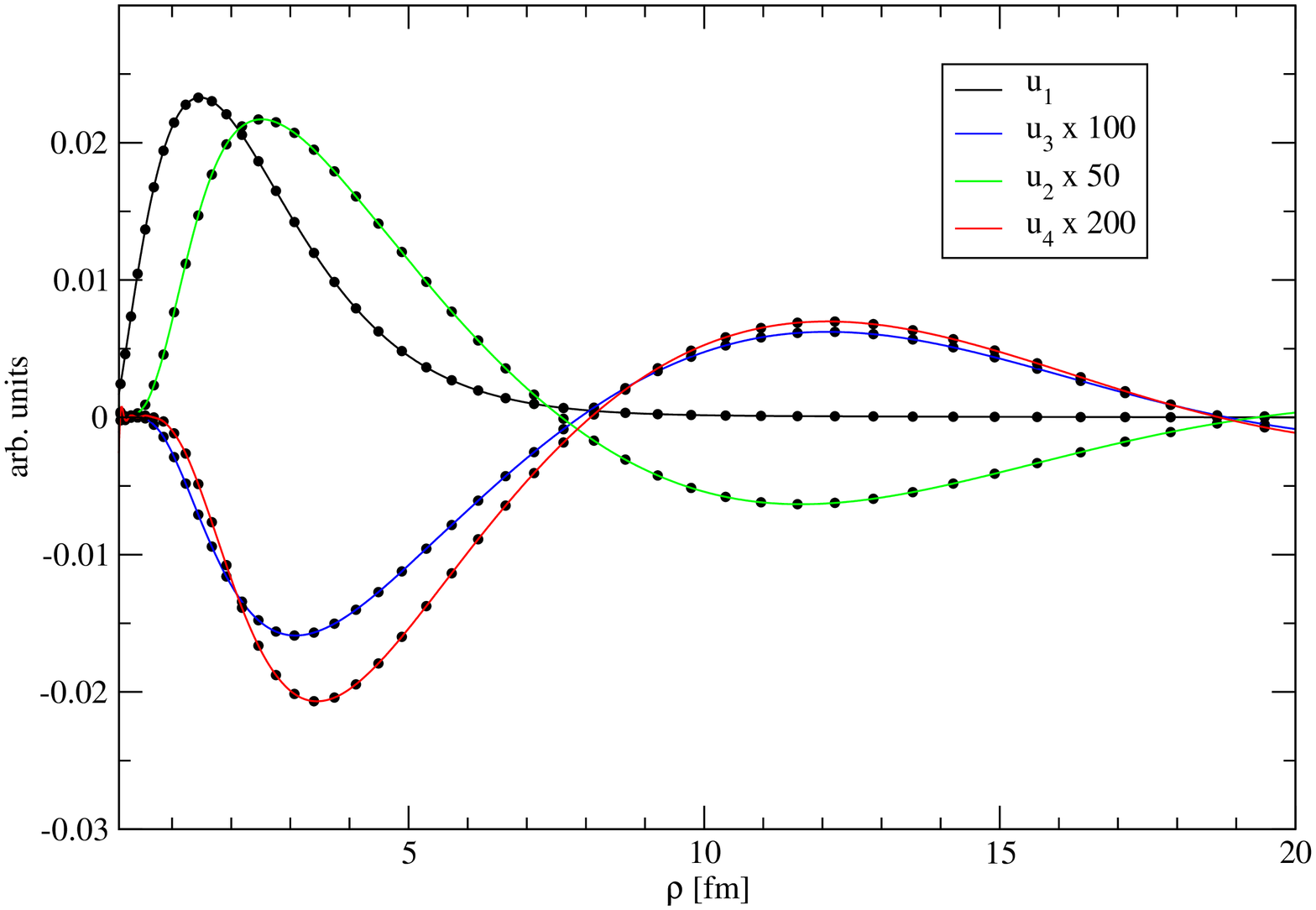}  &
\includegraphics[scale=0.20,angle=-90]{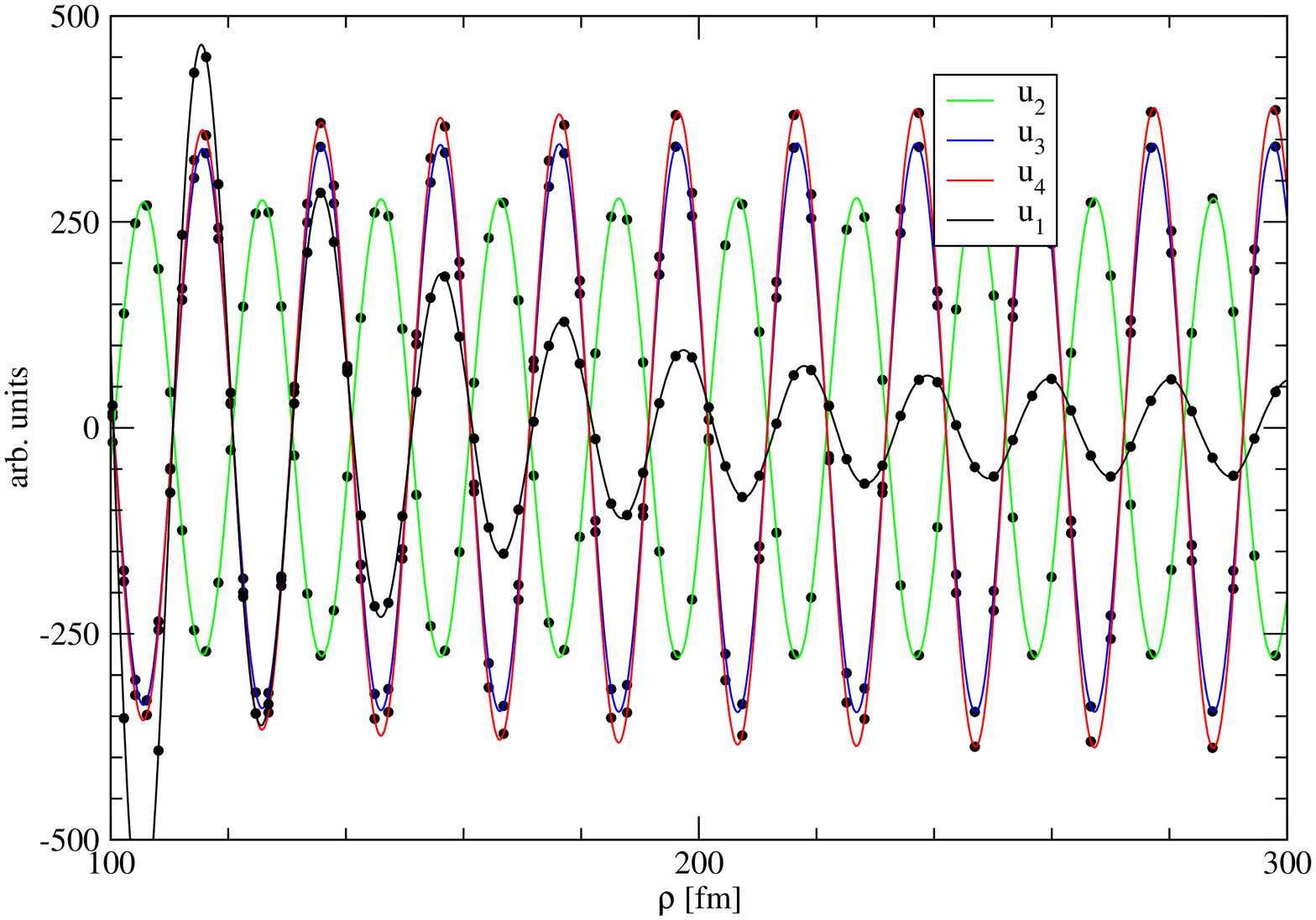}  
\end{tabular}
\caption{The functions $u_\nu(\rho)$ ($\nu=1,2,3,4$)
at $E=2.00$ MeV. The top panel shows the short-range region. 
Some of the functions were magnified by the factor shown in the legend. 
The circles represent the DVR amplitudes at the DVR grid points. 
%whereas the continuum, dashed, dotted and dot-dot-dash lines were 
%used to represent the functions $u_1$, $u_2$, $u_3$ and $u_4$, respectively.
The lower panel shows the long range region. In order to highlight the
 asymptotic behavior of each function, $u_1$ is multiplied by $\rho^{7/2}$, 
and $u_2$, $u_3$ and $u_4$ by $\rho^5$. In the bottom panel $u_1$ is also
 magnified by a factor 50000.}
\label{fig3}
\end{figure}

Fig. \ref{fig3} shows the functions 
$u_\nu(\rho)$ calculated with $N_A=4$, $N_{DVR}=300$ and $\rho_{max}=1200$ fm. 
 The dots indicate the DVR amplitudes at the 
DVR points, whereas the lines represent the $u_\nu(\rho)$
functions obtained by back-transforming 
to the original polynomial basis.
The top panel displays the short-range region ($0\le\rho\le 20$ fm), 
where the function $u_1$ is predominant.
The bottom panel shows a part of 
the long-range region ($100\le\rho\le 300$ fm).
Here the situation is drastically different, and the functions $u_2$, $u_3$ and $u_4$ have a much 
larger amplitude than $u_1$ (which is magnified by a factor $50000$). Also, in order 
to highlight the asymptotic behavior, $u_1$ is multiplied by $\rho^{7/2}$, and $u_2$,
$u_3$ and $u_4$ by $\rho^5$. The most striking feature are the oscillations present in  
all curves. This behavior is a consequence of the decomposition of the asymptotic
configuration in terms of HA functions.
This resulting peculiar long range behavior
is the cause of the very slow convergence of the phase-shift shown in Tables \ref{tab3a},
\ref{tab3}, \ref{tab4}. The behavior obtained for the curves $u_\nu$ is the one expected 
by the analytical expansion of the asymptotic terms indicated in 
eqs. (\ref{omexp1},\ref{omexpnu}). \\
\begin{figure}[h]
\includegraphics[scale=0.25,angle=-90]{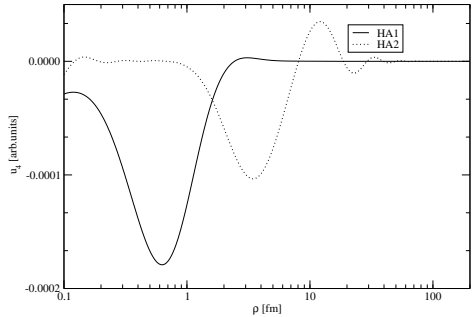}  
\caption{The function $u_4(\rho)$ (see eq.\refeq{adbasis}) obtained with the method HA1 (continuum line) 
and HA2 (dotted line). In the first case the function $u_4$ is short range, decaying 
exponentially with $\rho$, whereas in the second case it shows the 
long range oscillations.}
\label{fig4}
\end{figure}

Figure \ref{fig4} compares the hyperradial function $u_4(\rho)$ obtained with 
method HA1 and HA2. In particular it highlights as the former is short range 
and exponentially decaying with $\rho$, compared to the latter which is oscillating 
as indicated in eq. \refeq{omexp1}.
  
As mentioned in the Introduction, in Ref. \cite{fab1} the phase shift
for the potential $V_G$ has been calculated from eq.~\refeq{usys}
in the so-called uncoupled adiabatic approximation (UUA) retaining
one hyperradial function.
Namely, the following equation has been
solved:
\be
\left[ -\frac{\hbar^2}{2 m} T_\rho +U_1-E+ B_{11} \right] u_1(\rho) =0 
\label{usys1}
\ee
with the asymptotic condition $u_1(\rho)\rightarrow \sin(k\rho+\delta+3\pi/2)$
as $\rho\rightarrow\infty$. Besides the factor $3\pi/2$, this is equivalent to
the method HA2 given in the previous section taking into account one
HA function.\\
\begin{figure}[hbt]
\includegraphics[scale=0.25,angle=-90]{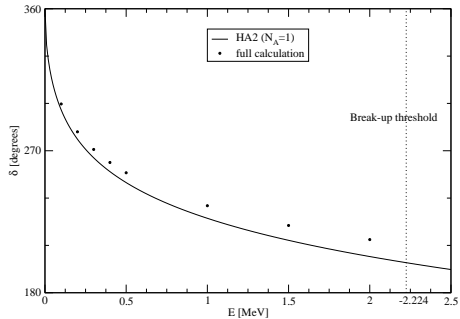}  
\caption{Elastic deuteron-nucleon phase-shift below three-body break-up (marked by the dotted line) 
for the $V_G$ potential. The full line corresponds to a calculation retaining one
HA basis element, and using the HA2 method. The dots correspond to the full calculation. }\label{fig5}
\end{figure}

In Figure \ref{fig5} we show the phase-shift $\delta(E)$.
The dots represent fully converged results obtained with 
the HA1 expansion, whereas the continuum line represent results obtained by 
including just one adiabatic function in the expansion HA2. 
It is possible to notice that the UUA provides a very good first order estimate 
of the phase-shift. However, 
the deviation from the complete expansion can be as big as 10\%.
Also notice that in Figure \ref{fig5}
the phase-shifts have been normalized
 so that $\delta(E=0)-\delta(E=\infty)=360$, as there 
are two bound trimer states. 

\section{Conclusions}

In this paper we have investigated the capability of the HA basis to describe 
scattering  states in a three-nucleon problem.
 The basis was generated from the hyperangular Hamiltonian by means of an 
expansion in HH functions. 
 We have shown the complete equivalence between the adiabatic basis
 generated
 using $N$ HH functions and the HH basis of dimension $N$. This
equivalence provides a useful benchmark when the convergence of the
 quantities of interest
 is studied in terms of the number $N_A$ of adiabatic functions.
For example, for bound
states it is well known that $N_A<<N$ suffices
 for the convergence of the binding energies.
 One goal of this paper was to investigate whether the same relation holds for
scattering states. In particular, we studied the convergence of the 
$L=0$ phase shift $\delta$ corresponding to a process in which a nucleon 
collides a deuteron at low energies in the state $S=3/2$. For this purpose 
we have used the MT-III potential. 

In the calculation of the phase shift using the 
HA basis we have followed two different procedures. They were both 
based on a decomposition of the scattering
wavefunction as a sum of two terms. One term
describes the configurations when the three particles are all close
 to each other and goes to zero as the interparticle distances increase.
 The second
term describes the asymptotic configurations and has been regularized
so that goes to zero as $y\rightarrow 0$. In the
first procedure the HA basis has been used to expand the short range part 
of the scattering wave function.
The second order estimate of the phase-shift has been obtained from
the Kohn variational principle. A similar approach has been used before
with the HH basis. Therefore, a detailed comparative analysis 
of the convergence patterns was possible. The conclusion is that the
number of basis elements needed to 
achieve a comparable level of convergence for the phase-shift 
is of the same order for the two bases,
 that is $N_A\approx N$, which is a surprising difference
with respect to what happens in bound state calculations. A possible
 explanation could be the following. In bound state calculations
the wavefunction expansion benefits from the initial 
optimization performed by constructing the HA basis. Conversely, in
scattering state calculations the solution of the linear system
of eq.\refeq{lsystem} requires a different short range behavior
in the HA basis elements due to the presence of the terms 
$\Omega^0_{ST}$ and $\Omega^1_{ST}$ in the short distance region.

The second procedure considered was based in a direct solution
of the system of equations for the hyperradial functions
given in eq.\refeq{usys}. This method however suffers from the following 
complications. The hyperradial boundary conditions to be imposed
 are those required to reconstruct the asymptotic
configuration given by the functions defined in eqs.(\ref{ha1r},\ref{ha1i}).
For very large values of $\rho$ the boundary conditions are simple
and are given by eq.~\refeq{asym} for the lowest function ($\nu=1$).
All other functions go to zero as $\rho\rightarrow\infty$.
This means that the solution of
the linear system has to be obtained over a very extended hyperradial grid.
 Moreover, the adiabatic potentials and functions have
to be accurately known in the grid. In the present work we
have solved partially the numerical difficulties associated to
the solution of eq.~\refeq{usys} introducing a variational DVR procedure.

From the present study we can conclude that  the use of the HA
basis in the description of scattering states is not as advantageous
as for bound states. The main drawback is that then number of basis elements
 required to reach convergence is not as low (in proportion) as in bound 
state calculations. Secondly, 
a number of numerical problems arise from the need of calculating 
the adiabatic curves and the associated basis elements at large distances.
Further studies to improve the description of scattering states
using the HA expansion are at present underway.


\begin{thebibliography}{10}

\bibitem{nielsen}
E.~Nielsen, D.~V. Fedorov, and A.~S. Jensen.
\newblock {\em J. Phys. B: At. Mol. Opt. Phys\/}, {\bf 31} (1998) 4085--4105.

\bibitem{bge00}
D.~Blume, C.~Greene, and B.~D. Esry.
\newblock {\em J. Chem. Phys.\/}, {\bf 113} (2000) 2145--2158.

\bibitem{dcf82}
Y.~Das, H.~Coelho, and M.~{Fabre de la Ripelle}.
\newblock {\em Phys. Rev. C\/}, {\bf 26} (1982) 2281.

\bibitem{bfr82}
J.~Ballot and M.~{Fabre de la Ripelle}.
\newblock {\em Phys. Rev. C\/}, {\bf 26} (1982) 2301.

\bibitem{ffs88}
M.~{Fabre de la Ripelle}, H.~Fiedeldey, and S.~Sofianos.
\newblock {\em Phys. Rev. C\/}, {\bf 38} (1988) 449.

\bibitem{fab1}
M.~{Fabre de la Ripelle}.
\newblock {\em Few-Body Systems\/}, {\bf 14} (1993) 1--24.

\bibitem{krv94}
A.~Kievsky, S.~Rosati, and M.~Viviani.
\newblock {\em Nucl. Phys. A\/}, {\bf 577} (1994) 511.

\bibitem{kvr97}
A.~Kievsky, M.~Viviani, and S.~Rosati.
\newblock {\em Phys. Rev. C\/}, {\bf 56} (1997) 2987.

\bibitem{cpf89}
C.~R. Chen, et~al.
\newblock {\em Phys. Rev. C\/}, {\bf 39} (1989) 1261.

\bibitem{pfg82}
G.~L. Payne, J.~L. Friar, and B.~F. Gibson.
\newblock {\em Phys. Rev. C\/}, {\bf 26} (1982) 1385.

\bibitem{gpk90}
V.~Gusev, et~al.
\newblock {\em Few-Body Syst.\/}, {\bf 9} (1990) 137--153.

\bibitem{hh2}
M.~{Fabre de la Ripelle}.
\newblock {\em Annals of physics\/}, {\bf 127} (1980) 62--125.

\bibitem{hh3}
M.~{Fabre de la Ripelle}.
\newblock {\em Annals of physics\/}, {\bf 147} (1983) 281--320.

\bibitem{abr}
M.~Abramovitz and I.~A. Stegun.
\newblock {\em Handbook of mathematical functions\/}.
\newblock Dover publications, New York,  (1970).

\bibitem{pb1}
P.~Barletta and A.~Kievsky.
\newblock {\em Phys. Rev. A\/}, {\bf A64} (2001) 042514.

\bibitem{k1}
A.~Kievsky.
\newblock {\em Nuclear Physics A\/}, {\bf 624} (1997) 125--139.

\bibitem{nota}
P.~Barletta and A.~Kievsky.
\newblock to be published.

\bibitem{dvr}
J.~C. Light and T.~{Carrington Jr.}
\newblock {\em Adv. Chem. Phys.\/}, {\bf 114} (2000) 263--310.

\bibitem{lbk04}
M.~Lombardi, P.~Barletta, and A.~Kievsky.
\newblock {\em Phys. Rev. A\/}, {\bf 70} (2004) 032503.

\bibitem{pap1}
P.~Barletta and A.~Kievsky.
\newblock {\em Few-Body Syst.\/},  (accepted for publication).

\end{thebibliography}
\end{document}